\documentclass[twocolumn]{aastex61}
\usepackage{graphicx}
\usepackage{subfigure}
\usepackage{amsmath}
\usepackage{epsfig}
\usepackage{lineno}

\newcommand{\imgdir}{./images_IV}
\newcommand{\refsec}[1]{Section~\ref{#1}}
\newcommand{\reffig}[1]{Figure~\ref{#1}}

\newcommand{\Illustris}{Illustris}
\newcommand{\TNG}{IllustrisTNG}
\newcommand{\autoGMM}{\texttt{auto-GMM}}

\shorttitle{Disk-, Bulge-, and Halo-dominated Galaxies}
\shortauthors{Du et al.}

\begin{document}
\title{The Evolutionary Pathways of Disk-, Bulge-, and Halo-dominated Galaxies}
\correspondingauthor{Min Du}
\email{dumin@pku.edu.cn}

\author{Min Du}
\affil{Kavli Institute for Astronomy and Astrophysics, Peking University, Beijing 100871, China}

\author{Luis C. Ho}
\affiliation{Kavli Institute for Astronomy and Astrophysics, Peking University, Beijing 100871, China}
\affiliation{Department of Astronomy, School of Physics, Peking University, Beijing 100871, China}

\author{Victor P. Debattista}
\affiliation{Jeremiah Horrocks Institute, University of Central Lancashire, Preston PR1 2HE, UK}

\author{Annalisa Pillepich}
\affiliation{Max-Planck-Institut f$\ddot{u}$r Astronomie, K$\ddot{o}$nigstuhl 17, D-69117 Heidelberg, Germany}

\author{Dylan Nelson}
\affiliation{Max-Planck-Institut f$\ddot{u}$r Astrophysik, Karl-Schwarzschild-Str. 1, 85741 Garching, Germany}

\author{Lars Hernquist}
\affiliation{Harvard--Smithsonian Center for Astrophysics, 60 Garden Street, Cambridge, MA 02138, USA}

\author{Rainer Weinberger}
\affiliation{Harvard--Smithsonian Center for Astrophysics, 60 Garden Street, Cambridge, MA 02138, USA}

\begin{abstract}
To break the degeneracy among galactic stellar components, we extract kinematic structures using the framework described in \citet{Du2019, Du2020}. For example, the concept of stellar halos is generalized to weakly-rotating structures that are composed of loosely bound stars, which can hence be associated to both disk and elliptical type morphologies. By applying this method to central galaxies with stellar mass $10^{10-11.5}\ M_\odot$ from the TNG50 simulation, we identify three broadly-defined types of galaxies: ones dominated by disk, by bulge, or by stellar halo structures. We then use the simulation to infer the underlying connection between the growth of structures and physical processes over cosmic time. Tracing galaxies back in time, we recognize three fundamental regimes: an early phase of evolution ($z\gtrsim2$), and internal and external (mainly mergers) processes that act at later times. We find that disk- and bulge-dominated galaxies are not significantly affected by mergers since $z\sim2$; the difference in their present-day structures originates from two distinct evolutionary pathways, extended vs. compact, that are likely determined by their parent dark matter halos; i.e., nature. On the other hand, slow rotator elliptical galaxies are typically halo-dominated, forming by external processes (e.g. mergers) in the later phase, i.e., nurture. This picture challenges the general idea that elliptical galaxies are the same objects as classical bulges. In observations, both bulge- and halo-dominated galaxies are likely to be classified as early-type galaxies with compact morphology and quiescent star formation. However, here we find them to have very different evolutionary histories.

\end{abstract}

\section{Introduction}

An accurate decomposition and classification of galaxies is required to uncover the causal link
between galaxy formation history and their properties. In observations, galaxies are generally 
decomposed by the limited information of their morphologies and kinematics. The presence of spirals,  
bulge-to-total ratio (e.g., the Hubble (1936) sequence), and rotation are widely used. 
These parameters permit us to infer certain aspects of a galaxy's evolution over cosmic time. However, 
the connection between these quantities and the galaxy formation history is still quite uncertain. For 
example, early-type galaxies (ETGs) exhibit featureless morphologies, 
notwithstanding the high frequency of bars and stellar rings in early-type S0 galaxies. 
For many years, this simple appearance was thought to reflect a straightforward formation via 
mergers that erases the diversity in both morphology and kinematics. More recent 
observations have shown that, while in many ways the structure of ETGs is intrinsically 
simple, there is a rich diversity of properties. It is well established now that ETGs can be separated into 
fast and slow rotators by their kinematics, thanks to the development of the integral-field unit (IFU) technique 
\citep[e.g.,][]{Emsellem2007, Emsellem2011, Cappellari2011a, Cappellari2011b}, which indicates very 
different formation and evolution histories. By applying the orbit-superposition Schwarzschild method 
\citep[e.g.,][]{Schwarzschild1979, Valluri2004, vandenBosch2008} to 
reconstruct stellar orbits, \citet{Zhu2018b} was able to make remarkable progress in decomposing
observed galaxies. \citet{Zhu2018a, Zhu2018b} showed that the kinematic structures they found 
exhibit several differences from the general expectation of morphological
decompositions. Kinematics help to break the degeneracy in the morphology of different stellar structures 
to a certain degree. However, it is still a very challenging, if not impossible, task to decompose 
galaxies accurately from observations, as galaxy formation histories are deeply 
encoded with complex physical processes, while the information observations can 
provide is limited. 

A significant degeneracy exists between bulges and stellar halos defined traditionally by morphological 
methods \citep{Du2020}, making several interpretations difficult. Within a $\Lambda$CDM hierarchical 
growth of structure scenario, there is no doubt that the formation of stellar halos is associated with 
mergers that disperse stars into large volumes or with the stellar stripping of low-mass orbiting satellites. 
Generally, however, bulges are also considered to be correlated with mergers 
\citep[e.g.][]{Toomre1977, Aguerri2001, Hopkins2010, Wellons2015}, i.e., external processes. 
In fact, a variety  of internal processes that conspire to produce gas-rich inflows are possibly also 
important in bulge formation \citep{Dekel&Burkert2014, Zolotov2015, Tacchella2016b}. Such processes 
include disc instabilities, clump migration, and misaligned gas streams 
\citep[e.g.][]{Dekel2009, Parry2009, Bournaud2011, Sales2012, Ceverino2015, Wellons2015, Park2019, Guo2020}, 
which are closely associated with the underlying dark matter halos that galaxies inhabit. In such a picture, galaxy 
sizes and angular momenta are expected to be controlled by halo angular momenta, as gas cools out of 
gaseous halos that are initially coupled with their parent dark matter haloes 
\citep{Mo1998, Bullock2001, Zolotov2015}. However, whether the angular momentum can be conserved 
sufficiently is still under debate. \citet{Jiang2019} did not find a clear correlation for 
galaxies from zoom-in hydro-cosmological simulations, perhaps because of the change of angular momentum 
when cold streams fall into the inner regions of halos \citep{Danovich2015}.

A complete galaxy formation theory can almost only be achieved with numerical 
simulations and, in particular, with cosmological simulations that self-consistently 
evolve the dark matter and baryonic components of the Universe from cosmologically-motivated 
initial conditions. In such simulations, galaxies naturally emerge in a great diversity under 
the influence of internal and external processes \citep{Vogelsberger2014a, Schaye2015}. In 
the first few billion years of cosmic evolution, young galaxies form from efficient gas 
accretion and then rapid star formation (SF), i.e., the epoch known as ``cosmic noon'', at $z\sim 2-3$.
Because of the difference in the early-phase evolution, a bimodality in galaxy type can 
begin to occur \citep{Dekel2009}, which will possibly lead to long-lasting differences at 
subsequent cosmic epochs. In such later phases, galaxies move into a secular evolution period 
driven by internal processes in the cases of no significant merger activity. Gas and stellar 
velocity dispersions decrease toward low redshifts with the decrease of star formation and 
the increase of the galaxy potential well 
\citep[e.g.][]{Law2009, Daddi2010, Geach2011, Genzel2011, Swinbank2012, Dessauges-Zavadsky2015, Girard2018}. 
Rich structures, e.g., bars, rings, pseudo-bulges \citep[reviewed by][]{Kormendy&Kennicutt2004}, generated 
largely by internal instabilities, partially account for the rich galaxy diversity. 

However, though mergers are progressively rarer at lower redshifts, they can dramatically 
change the morphology and kinematics of galaxies when they happen, especially major ones. 
It is well known that dissipationless dry minor/major mergers can disrupt galaxy spin, generating 
ETGs \citep{Khochfar&Silk2006, Naab2006, vanderWel2009, Bezanson2009}. It has been suggested that the 
cumulative effect of many dry minor mergers can explain the size growth of ETGs from $z=2$ to the 
present via the buildup of a diffuse envelope. However, recent analyses on the 
Illustris(TNG) and EAGLE simulations do not see a clear cumulative effect of minor mergers 
\citep{Penoyre2017, Lagos2018, Pulsoni2021}. Instead, dry major mergers generally lead to the formation of massive 
slow-rotating ETGs, especially for central ones \citep{Lagos2018, Pulsoni2021}. Even more interestingly,  
$\approx 30\%$ of the ETGs in EAGLE have not had any mergers with mass ratios $\geq 0.1$ 
during their past 10 Gyr. This fraction is smaller in more massive galaxies. Similarly, \citet{Penoyre2017} and \citet{Pulsoni2021}
also suggested that low-mass ($M_{\rm s} < 10^{11} M_\odot$) ETGs have a very different assembly 
history from high-mass ones. 

In recent years, significant progress has been made in reproducing realistic galaxy morphologies particularly in large-volume 
hydrodynamical simulations like \Illustris\ \citep{Genel2014, Vogelsberger2014a, Vogelsberger2014b, Nelson2015, Sijacki2015},  
EAGLE \citep{Schaye2015, Crain2015}, and Horizon-AGN \citep{Dubois2016}. The \TNG\ simulations 
\citep{Nelson2018a, Nelson2019a, Naiman2018, Marinacci2018, Pillepich2018b, Pillepich2019, Springel2018}
is the advanced version of \Illustris. It can reproduce galaxies that successfully emulate real galaxies in many 
aspects, thanks to a well-designed galaxy physics model \citep{Weinberger2017, Pillepich2018a}. In simulations, we are 
able to extract intrinsic structures in a physical way, as well as to track their formation processes and evolutionary 
histories. This provides insights into the formation history of real galaxies that display a great diversity. 

Understanding the evolution of galaxies in numerical simulations is required to help in recovering 
the comparable evolution of real galaxies. As a first step in this process, we developed a fully automatic 
Gaussian mixture model, called \autoGMM\, that can decompose simulated galaxies in a non-parametric, accurate, 
and efficient way \citep{Du2019}. This method takes full use of the 6D information of the position and velocity 
for every star (i.e. stellar particle). By applying \autoGMM\ to about 4000 disk galaxies from the TNG100 run 
of the \TNG\ suite, we uncovered rich kinematic structures that statistically cluster well in the 3D space 
of structural kinematic moments \citep{Du2020}. The structural kinematic moments are composed of dimensionless 
binding energy, circularity parameter, and non-azimuthal angular momentum that quantify the compactness, 
circular rotation in the disk aligned with the global spin, and the mis-aligned rotation, respectively, of 
each structure. We define the structures with strong to moderate rotation as cold and 
warm disks, respectively. Spheroidal structures dominated by random motions are classified as bulges or 
stellar halos, depending on how tightly bound they are. 
\citet{Du2020} suggested that the morphological decomposition widely used in observations can 
barely represent kinematic structures found in the simulations that are likely corresponding to
intrinsic structures. We showed that morphologically-derived bulges are largely composites of 
kinematic bulges, stellar halos and even disky bulges in their inner regions. This may lead to serious biases in the 
physical interpretation of such structures. Our kinematic decomposition method, thus, has potential to gain 
great insights into the evolutionary histories of real galaxies.

In a series of works (including this one), we aim to understand the formation history of galaxies using a 
framework based on kinematically-derived intrinsic structures. In this paper, we apply \autoGMM\ to the TNG50 simulation.
This enables us to study realistically-simulated galaxies in unprecedented detail and statistics. We regard all 
processes at $z\gtrsim2$, that are quite chaotic, as the early-phase evolution of galaxies. At $z\lesssim2$, galaxy 
evolution can be influenced by internal and external (mainly but not exclusively mergers) processes. This description 
of galaxy formation is similar to the `two-phase' picture: an early phase of dissipative collapse and a later 
phase of dissipationless mergers \citep{Oser2010}. However, here we separate the physical processes of the later 
phase into internal/in-situ and external/ex-situ ones. The rich diversity in kinematic structures will be 
interpreted in the context of these three regimes. 

The paper is organized as follows. Sections \ref{sec:sim} and \ref{sec:method} introduce the sample selection 
and our kinematic decomposition method, respectively. Some basic properties of galaxies with various kinematic 
structures are shown in \refsec{sec:relation}. Galaxies are quantified, even classified, by the mass 
fractions of their kinematic structures in \refsec{sec:kinemclass}.  In Sections \ref{sec:bulgeanddisk} and 
\ref{sec:halo}, we then study the formation history of three kinds of typical galaxies that are dominated by 
disk, bulge, and stellar halo structures, respectively. \refsec{sec:discussion} discusses the results, 
then the main conclusions are summarized in \refsec{sec:conclusion}.

\section{The TNG50 simulation}
\label{sec:sim}

The \TNG\ suite comprises three runs using different simulation volumes and resolutions, namely TNG50, TNG100, and TNG300. 
The simulations are run with gravo-magnetohydrodynamics (MHD) and incorporate a comprehensive galaxy model 
\citep[see][for details]{Weinberger2017, Pillepich2018a}. This study uses the smallest volume run, TNG50, which provides a 
large enough number of galaxies for statistical analyses and a ``zoom''-like resolution in which stellar particles have mass $\sim 10^4 M_\odot$. The TNG50 data is now publicly 
available at \href{https://www.tng-project.org}{https://www.tng-project.org}. For comparison, we also include some 
results from TNG100 in Appendix \ref{appendix}. First results from this simulation focusing on galactic outflows and the 
formation of rotationally supported disks are presented in \citet{Nelson2019} and \citet{Pillepich2019}. 
TNG50 includes $2 \times 2160^3$ initial resolution elements in a $\sim 50$ comoving Mpc box, corresponding to a 
baryon mass resolution of $8.5 \times 10^4 M_\odot$ with a gravitational softening length for stars of 
about $0.3$ kpc at $z = 0$. Dark matter is resolved with particles of mass $4.5 \times 10^5 M_\odot$. Meanwhile, 
the minimum gas softening reaches 74 comoving parsec. TNG50 thus has roughly 15 times better mass resolution, and 2.5 
times better spatial resolution, than TNG100 \citep[also publicly available, see][]{Nelson2019a}. 
 
As in other cosmological simulations, also in TNG50 galactic outflows driven by feedback from both supernovae 
and supermassive black holes are key ingredients in generating galaxies with realistic morphologies over a broad mass 
range \citep{Nelson2019}. The unprecedented resolution allows us to study a large sample of galaxies with the details 
that were previously achieved only in zoom-in simulations. \citet{Pillepich2019} showed that star forming galaxies in 
TNG50 have a typical thickness of a few hundred parsecs, in much better agreement with observations than TNG100 at 
lower resolution. Both the thickness and kinematics of galaxies above $M_{\rm s} =10^9 M_\odot$ are now reasonably 
converged. Moreover, TNG50 can resolve many physical processes down to small scales, for instance, cold gas clouds 
in the circumgalactic medium (CGM) with sizes $\sim$ a few hundred parsecs that are stabilized by magnetic fields 
\citep{Nelson2020}, fine-grained galaxy stellar morphological structures \citep{Zanisi2021}, stellar halo mocks 
similar to Dragonfly galaxies \citep{Merritt2020}, and metallicity gradients \citep{Hemler2020}.

TNG galaxies are identified and characterised with the Friends-of-Friends \citep[FoF][]{Davis1985} and {\sc SUBFIND} 
\citep{Springel2001} algorithms. Resolution elements (gas, stars, dark matter, and black holes) belonging to an 
individual galaxy are gravitationally bound to its host subhalo. In this work, we focus on TNG50 galaxies with total 
stellar mass of $M_{\rm s} = 10^{10}-10^{11.5} M_\odot$, including both spirals and ellipticals. We do not adjust the galaxy stellar masses to account for possible resolution effects \citep{Pillepich2018b, Engler2021}. There are 873 
galaxies satisfying this criterion at $z=0$ in TNG50. Our main conclusions are based on central galaxies to avoid 
possible environmental effects \citep[e.g.][]{Joshi2020, Engler2021}, resulting in a sample of 541 galaxies. 

\section{Extracting kinematic structures: methodology}
\label{sec:method}

We identify kinematic structures in galaxies from TNG50 with the framework introduced in \citet{Du2019, Du2020}. We 
here give only a brief overview of the method: all that follows is applied exclusively to the stellar component of galaxies.

The first step is to physically characterize stars in the phase space of any individual galaxy. In this series of works, 
we use the kinematic phase space comprised of the circularity parameter 
$\epsilon=j_z/j_c(e)$ \citep{Abadi2003b}, the non-azimuthal angular momentum $j_p/j_c(e)$, and the binding energy normalized 
by the minimum value $e/|e|_{\rm max}$, as proposed by \citet{Domenech-Moral2012}, of each stellar particle. Thus, 
$j_z/j_c$ and $j_p/j_c$ are physical parameters that quantify the aligned and misaligned rotation with the overall 
angular momentum, respectively, and $e/|e|_{\rm max}$ describes how tightly bound a stellar particle is. Part of the code 
from \citet{Obreja2018a} is used to build the kinematic phase space for gravitationally-bound stars to a galaxy.

Then, an automatic Gaussian mixture model\footnote{Gaussian mixture models (GMM) are unsupervised machine 
learning algorithms that are widely used to model discrete points with multidimensional Gaussian distributions. 
Here we use the {\tt GaussianMixture} module in the \texttt{PYTHON} {\tt scikit-learn}.} \autoGMM, is used to model the 
kinematic phase space. Stars are classified into multiple Gaussian components with ``soft'' probabilistic 
assignment. \citet{Obreja2018a, Obreja2019} made the first attempt to extract classical/pseudo-bulges and stellar 
halos by applying GMM to this parameter space, switching from the $K$-means clustering algorithm used in 
\citet{Domenech-Moral2012}. The number of Gaussian components was chosen artificially in these methods, which further 
leads to human bias in the identification of structures. As recommended by \citet{Du2019}, \autoGMM\ 
allows the number of Gaussian components to 
be determined automatically by setting the modified Bayesian information criterion $\Delta{\rm BIC}<0.1$, 
which corresponds to a Bayes factor $0.95-1$ with respect to the ideal model using numerous Gaussian 
components. In this case, we consider that this model performs equally well as the ideal model 
in a statistical sense. Generally, 4-9 prominent Gaussian components in the kinematic 
phase space will be found for modelling any individual galaxy properly. The number of Gaussian 
components is inferred directly from the data. For each component, its kinematics can be quantified by the 
mass-weighted average values of $j_z/j_c$, $j_p/j_c$, and $e/|e|_{\rm max}$, defined as 
$\langle j_z/j_c \rangle$, $\langle j_p/j_c \rangle$, and $\langle e/|e|_{\rm max} \rangle$. This method 
not only successfully avoids overfitting due to the use of too many components, but also minimizes 
the possibility of human bias, which makes it possible to identify intrinsic structures in galaxies.

Finally, the intrinsic structures of galaxies are then objectively inferred from statistical results. 
In \citet{Du2020}, via stacking all components together in thousands of disk galaxies from TNG100, we 
found that the stellar components also cluster in the kinematic-moment space composed of $\langle j_z/j_c \rangle$, 
$\langle j_p/j_c \rangle$, and $\langle e/|e|_{\rm max} \rangle$. We have thus identified the following 
useful classification: 
\begin{itemize}
\item clusters of stars having strong ($\langle j_z/j_c \rangle \geq 0.85$) to moderate ($0.5 \leq \langle j_z/j_c \rangle < 0.85$) rotation are defined as cold and warm disks, respectively; 
\item clusters of stars dominated by random motions ($\langle j_z/j_c \rangle < 0.5$) and tightly bound ($\langle e/|e|_{\rm max} \rangle \leq -0.75$) are classified as bulges;
\item clusters of stars dominated by random motions ($\langle j_z/j_c \rangle < 0.5$) but that are loosely bound ($-0.75<\langle e/|e|_{\rm max} \rangle$) are defined as stellar halos.
\end{itemize}
All Gaussian components in the kinematic phase space are thus exclusively classified into cold disk, warm disk, bulge, and halo 
stellar structures. The overall disk and spheroidal structures are obtained by summing stars of the cold+warm disks and of the bulge+halo, 
respectively. Such criteria have been heuristically inferred from the statistical analysis on the disk galaxies from the TNG100 simulation, 
as presented in \citet{Du2020}. This classification method 
is the simplest and physically clearest classification of galaxy intrinsic structures, referred to as classification 1 in \citet{Du2020}. 

It is worth emphasizing that kinematically-defined disky structures in such a classification generally do not 
follow a simple exponential profile, in contrast with what has been typically and widely used in morphological 
decompositions \citep{Du2020}. The overall kinematic disks obtained by summing stars of cold and warm 
disks commonly have extra mass in their central regions, where auto-GMM can further isolate disky bulges that 
have bulge-like compact morphology ($\langle e/|e|_{\rm max} \rangle \leq -0.65$) \citep[][see Figures 6 and 7]{Du2020} but moderate rotation 
($\langle j_z/j_c \rangle \geq 0.5$), as defined in classification 2 of \citet{Du2020}. Disky bulges are rotation-dominated 
as well as centrally-concentrated structures. They generally lead to a clear 
deviation from exponential profiles in the central regions of overall kinematically-derived disks. It is true that this 
deviation will be reduced if we use a larger $\langle j_z/j_c \rangle$ threshold for kinematic disks, one that excludes disky bulges and warm disks. 
However, both of these are apparently disky and rotation-dominated structures with $\langle j_z/j_c \rangle \geq 0.5$ 
(see examples in Figure 5 of \citet{Du2020} and in Section \ref{sec:bulgeanddisk} of this paper). They are 
also likely to be formed by in situ processes, while they contaminate the face-on surface density of 
morphologically-decomposed bulges by $\sim30\%$ at $R<2$ kpc \citep[see Figure 14 of ][]{Du2020} in disk galaxies. 
This phenomenon stresses again the importance of an accurate decomposition of galaxies using their kinematics. 
Cold disks, on the other hand, are often truncated in their inner regions 
(see also arguments based on observations in \citet{Zhu2018b} and \citet{Breda2020})\footnote{The relation 
between such inner truncation in kinematic cold disks and the inner break found in purely photometric decomposition 
\citep{Gao&Ho2017} is unclear.}. In this paper, we adopt throughout the simpler classification 1 (cold and warm disks, 
bulges, stellar halos): using a more complex methodology such as classification 2 does not affect our results in this paper. 

Following the three steps above, we decompose all galaxies in the TNG50 sample into kinematic stellar cold disk, 
warm disk, bulge, and halo structures\footnote{The mass fractions of kinematic structures are 
publicly released at \href{https://www.tng-project.org/data}{www.tng-project.org/data}.} 
which qualitatively correspond to thin disks, thick disks + pseudo bulges, 
classical bulges, and stellar halos (also diffuse envelopes in elliptical galaxies) in observations, 
respectively. They will be used in the subsequent analysis. All stars bound to 
the galaxy are counted, in order to measure mass fractions of these kinematic structures accurately. 
It is worth mentioning that this mass fraction cannot be directly compared with observations where, generally, 
stellar light is probed only out to a few effective radii or less.

\begin{figure*}[htbp]
\begin{center}
\includegraphics[width=0.8\textwidth]{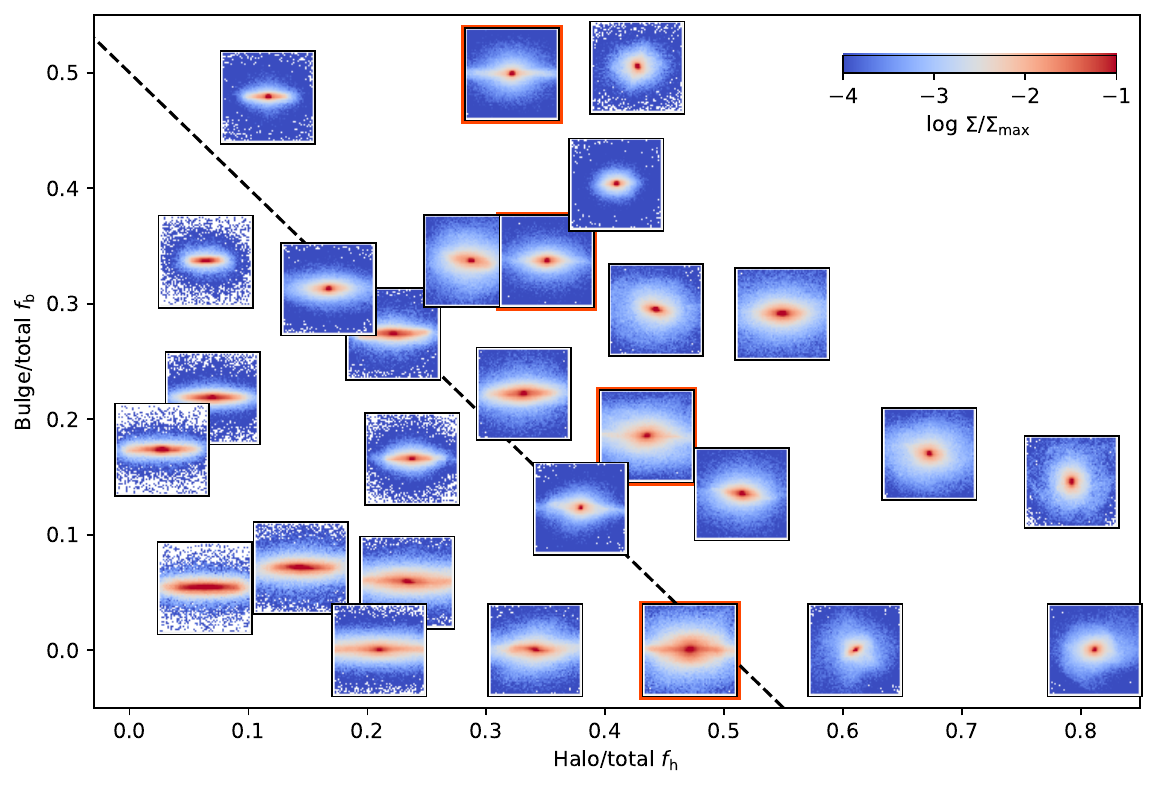}
\caption{Edge-on views of a randomly selected sample of $z=0$ TNG50 central galaxies in the mass range $M_{\rm s} = 10^{10.5}-10^{11} M_\odot$, in the bulge-to-total vs. stellar halo-to-total stellar mass fraction plane. For each galaxy, the edge-on surface density maps are shown in a region of $40\times 40$ kpc. The surface density is normalized by the maximum value for each galaxy. The dashed line marks the position where the mass fraction of spheroidal components is equal to 0.5, i.e., $f_{\rm b} + f_{\rm h} = 0.5$: it can be used to separate disk galaxies from elliptical ones. A massive central concentration commonly exists in galaxies with massive bulges, while the galaxies with massive halos are generally surrounded by diffuse envelopes. The red squares mark four TNG50 analogues of the Sombrero Galaxy.}
\label{fig:example_dens}
\includegraphics[width=0.8\textwidth]{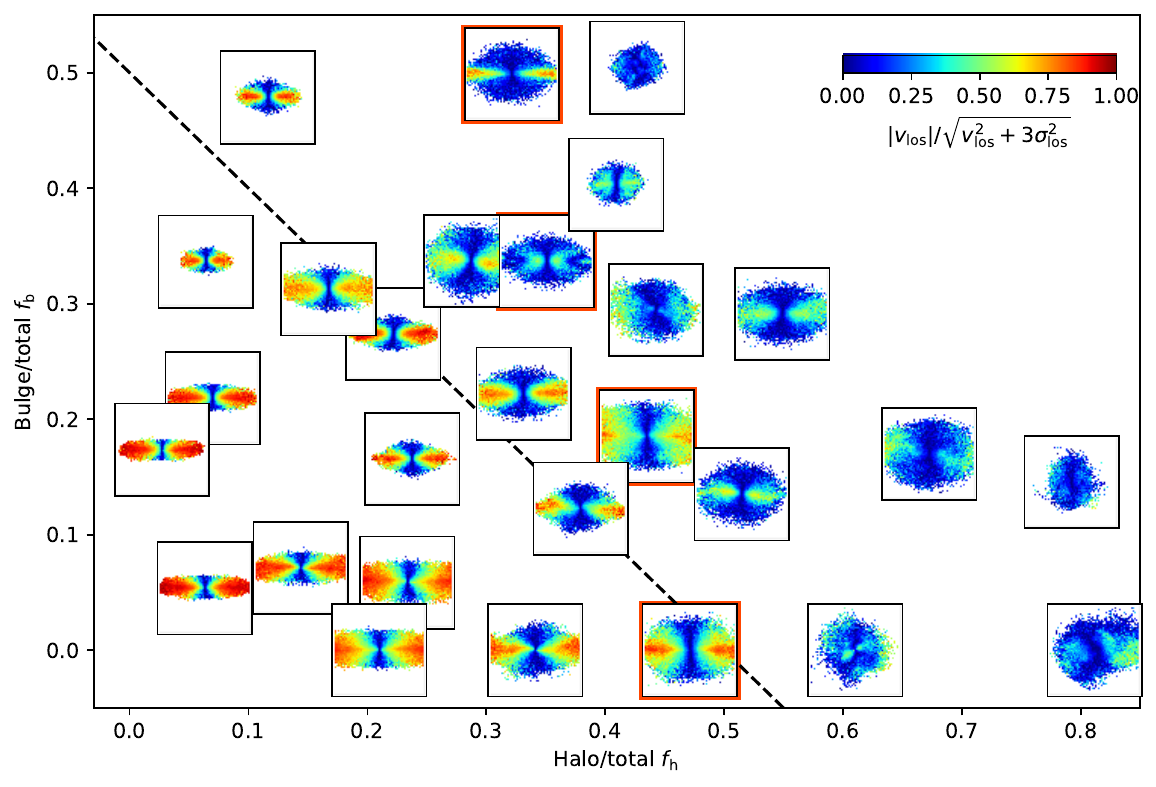}
\caption{As in \reffig{fig:example_dens} but for the relative importance of rotation $\vert v_{\rm los} \vert/\sqrt{v_{\rm los} + 3\sigma_{\rm los}}$, estimated from the edge-on view of the same TNG50 galaxies, where $v_{\rm los}$ and $\sigma_{\rm los}$ are the mean velocity and velocity dispersion in the line-of-sight view, respectively. The rotation becomes weaker and weaker with increasing spheroidal fraction towards the top-right corner.}
\label{fig:example_vos}
\end{center}
\end{figure*}

\begin{figure*}[htbp]
\begin{center}
\includegraphics[width=0.8\textwidth]{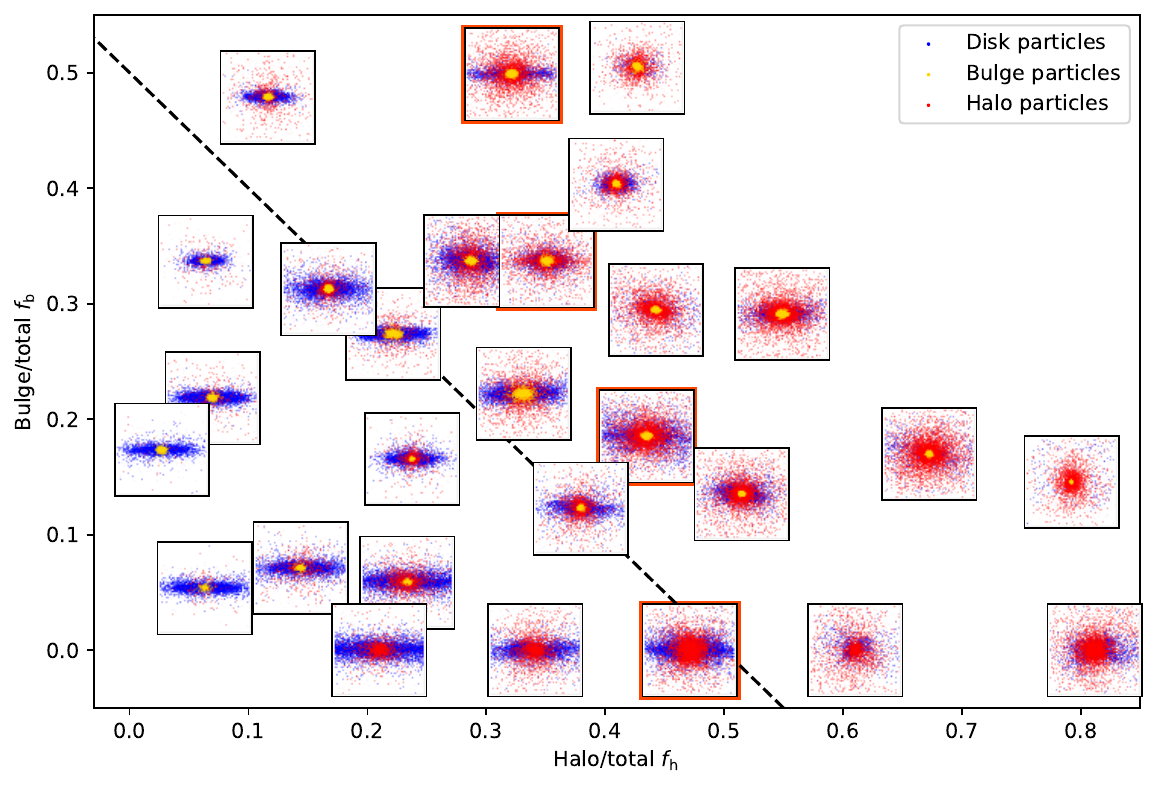}
\caption{The edge-on spatial distributions of disk, bulge, and stellar halo particles selected by our kinematic decomposition method for the same TNG50 galaxies as in Figures \ref{fig:example_dens} and \ref{fig:example_vos}. For each galaxy, $10^5$ stellar particles are selected randomly. Bulges are generally concentrated in the central regions of galaxies, while halos typically follow a diffuse distribution that extends from the center to an extended envelope. It is worth emphasizing that there is a severe degeneracy between bulges and halos in the central regions of galaxies. Here we plot bulge particles last in order to make them more visually prominent.}
\label{fig:example_p}
\includegraphics[width=0.8\textwidth]{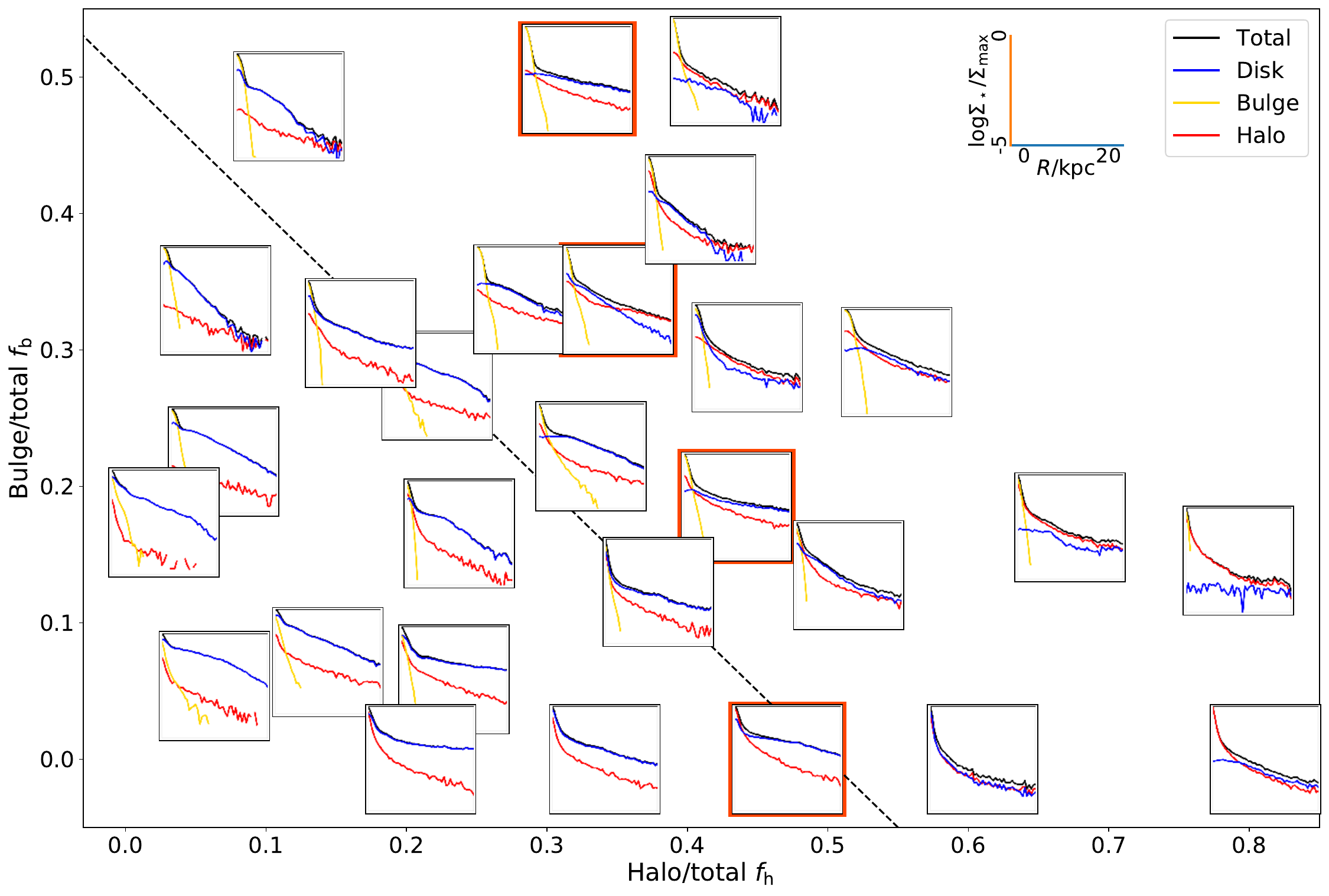}
\caption{Distribution of the normalized surface density profiles in the midplane for the same TNG50 galaxies as 
Figures \ref{fig:example_dens}, \ref{fig:example_vos}, and \ref{fig:example_p}. The results of all stars, and those of kinematic disk, bulge, and halo structures are shown in black, blue, yellow, and red, respectively. For each panel, the $x$- $y$-axes represent $R$/kpc and log$\Sigma_\star/\Sigma_{\rm max}$, respectively, covering the 
range of [0, 20] kpc and [-5, 0] (top-right corner). $\Sigma_{\rm max}$ is the maximum value of 
$\Sigma_\star$. Neither bulges nor halos are necessarily the direct counterparts of morphological 
bulges described by the S$\acute{e}$rsic function.} 
\label{fig:profile}
\end{center}
\end{figure*}

\section{Relation between kinematic structures and global properties}
\label{sec:relation}

\subsection{A physical definition of bulges and stellar halos}

Cosmologically-motivated models suggest that stars tend to conserve their binding energy during galaxy mergers. 
Ex-situ stars, thus, can be loosely bound and can hence populate galaxies in a broad range of 
galactocentric distances \citep[e.g.][]{Barnes1988, Hopkins2009, Amorisco2017}. The constituent stars of 
stellar halos are fossil records of the hierarchical merging process 
\citep[e.g.][]{Deason2016, DSouza&Bell2018, Monachesi2019}: mergers with larger 
satellites produce more massive, higher-metallicity stellar halos, and can thus reproduce the recently observed stellar halo metallicity-mass relation \citep[discovered by an {\it HST} imaging survey of nearby galaxies, GHOSTS,][]{Harmsen2017}.
Studies of the hierarchical growth of structures have reached similar conclusions using 
large-scale, hydrodynamic cosmological simulations \citep[e.g. \Illustris][]{Pillepich2014, Rodriguez-Gomez2016, Pop2018}. 

It is often argued that massive, compact classical bulges formed at early cosmic epochs via various pathways such as early gas-rich accretions, violent disk instabilities, or misaligned inflows of gas. Such classical bulges, thus, are largely composed of stars formed in-situ and characterized by low binding energy. 
\citet{Bell2017} showed that galaxies with massive classical bulges have diverse merger 
histories, and no clear correlation with properties of the stellar halos has been found. It is, thus, plausible that bulges are indeed dominated by in-situ chaotic processes. The classical conception that bulges are produced by mergers may therefore not hold in all cases.

Figures \ref{fig:example_dens} and \ref{fig:example_vos} show the distributions of the morphology 
and relative importance of rotation for a selection of TNG50 galaxies when viewed edge-on. 
They are characterized by the mass fraction of kinematic bulge ($f_{\rm b}$, $y$-axis) and halo 
($f_{\rm h}$, $x$-axis) derived by \autoGMM. We normalize the surface density map of each galaxy by its maximum value to gain equal contrast for galaxies with different stellar masses. Obviously, disk galaxies have strong rotation, thus located at the bottom-left corner. Elliptical galaxies mainly lie above the dashed line where the overall mass fraction of spheroids $f_{\rm sph} = f_{\rm b}+f_{\rm h}$ is larger than 0.5. 
However, galaxies with massive bulges seem to be not clearly distinguishable from those having massive halos in observations, even when taking kinematics into account. This issue is the more serious the lower the inclination. Therefore, a severe degeneracy exists in classical morphological decompositions of bulge vs. halo stars, even though the central massive concentration indeed becomes more prominent with the increase of bulge mass fraction. 

\begin{figure*}
\begin{center}
\includegraphics[width=1\textwidth]{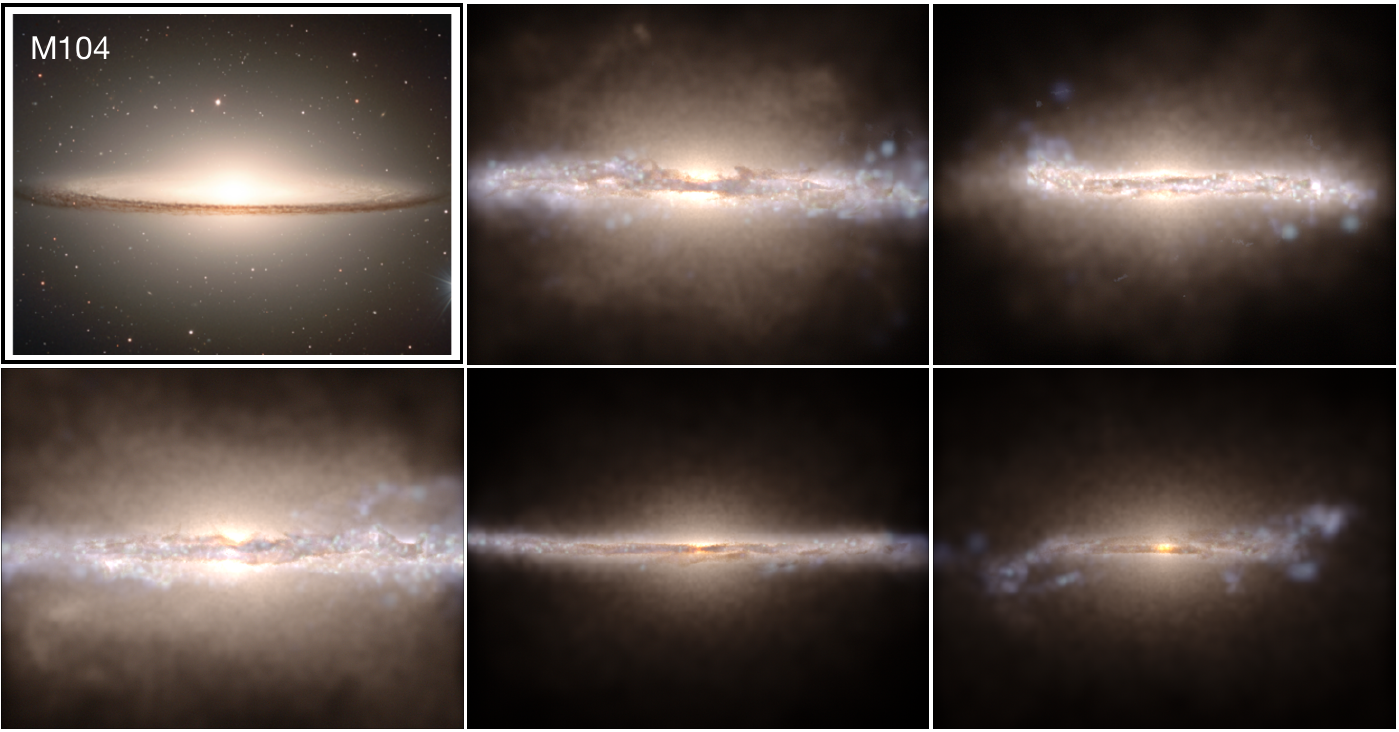}
\caption{The Sombrero Galaxy (M104), top left panel, and analogues from the TNG50 simulation. M104 is an example of a galaxy with a large, classical central ``bulge'' that is possibly a stellar halo according to our kinematic definition. The top left is a composite image of V, R, I-bands that was obtained with the FORS1 multi-mode instrument at VLT ANTU [\href{https://www.eso.org/public/images/eso0007a/}{ESO}] in a field of view of about 30 kpc. The other panels showcase idealized synthetic images of TNG50 galaxies (using HST/ACS  F435W, F606W and F775W filters) generated with the radiative transfer code {\sc SKIRT} as in \citet{Rodriguez-Gomez2019}. Unlike the real Sombrero galaxy, the simulated objects are seen perfectly edge-on and across a larger field of view of about 70 kpc. The kinematically-defined stellar halos of these galaxies indeed occupy 35-50 per cent of their total stellar masses.} 
\label{fig:M104}
\end{center}
\end{figure*}
As discussed in \refsec{sec:method}, the stars of bulges and stellar halos defined by our kinematic 
method are separated by their binding energies, which is consistent with our physical expectation. 
Both bulges and halos have similarly weak rotation; however, as shown by the spatial distribution of 
their stellar particles in \reffig{fig:example_p} and 1D surface density profiles in \reffig{fig:profile}, 
bulge stars (yellow) are tightly bound around the galactic central regions. Halo stars (red) are loosely 
bound, comprising the diffuse envelopes. Halo stars, that move on highly elliptical orbits, are able to pass 
through the central regions that are dominated by bulge stars. The half-mass radii of bulges are generally 
less than 2 kpc, while those of the stellar halos vary in a broad range of 2-10 kpc.

The fact that stellar halos approximately follow the S$\acute{e}$rsic function, see \reffig{fig:profile}, 
may induce a serious difficulty in making accurate morphological decompositions of galaxies, and hence 
in advancing interpretations for their formation processes. In order to illustrate this issue clearly, 
we take the famous Sombrero Galaxy (M104/NGC 4594, see \reffig{fig:M104}) as an example. The Sombrero 
Galaxy is regarded as one of the most unusual galaxy having a disk embedded in an extremely large 
``bulge''. \citet{Gadotti&Sanchez-Janssen2012} argued 
that the bulge mass fraction can be reduced from $77\%$ to $<10\%$ in the Sombrero Galaxy 
if an outer spheroidal component, i.e., a stellar halo, is considered. We highlight four Sombrero 
visual analogues with red squares in Figures \ref{fig:example_dens} and \ref{fig:example_vos}. 
As we can see in Figures \ref{fig:example_p} and \ref{fig:profile}, the huge ``bulges'' of Sombrero 
analogues are largely contributed by kinematic halos. The mass fraction of their bulges can vary 
from 0 to 0.5. Idealized synthetic images of Sombrero-like galaxies from TNG50 are shown too in 
Figure~\ref{fig:M104}, following the procedure described in \citet{Rodriguez-Gomez2019}.

auto-GMM allows us to break the degeneracy between bulges and halos even in the central regions 
of galaxies. In this picture, kinematic bulges qualitatively correspond to classical bulges. 
Normal elliptical galaxies are largely dominated by kinematic stellar halos, which challenges the 
general idea that elliptical galaxies (i.e., slow rotator ETGs) are the same objects as classical bulges in disk galaxies, 
obeying the Kormendy relation \citep[e.g.][]{Kormendy1977, Gadotti2009} and the $M_{\rm bh}-\sigma_{\rm s}$ 
relation \citep[e.g.][]{Kormendy&Ho2013}. About 52 per cent of our central elliptical galaxies, 
that are selected by the spheroidal mass fraction $f_{\rm sph} \geq 0.5$, are halo-dominated galaxies. It is worth 
emphasizing that although stellar halos have generally much lower surface density than other structures 
on the midplane, their overall mass fractions can be large due to their wide extent reaching tens, 
if not hundreds, of kpc distance. 

\begin{figure*}[htbp]
\begin{center}
\includegraphics[width=1.\textwidth]{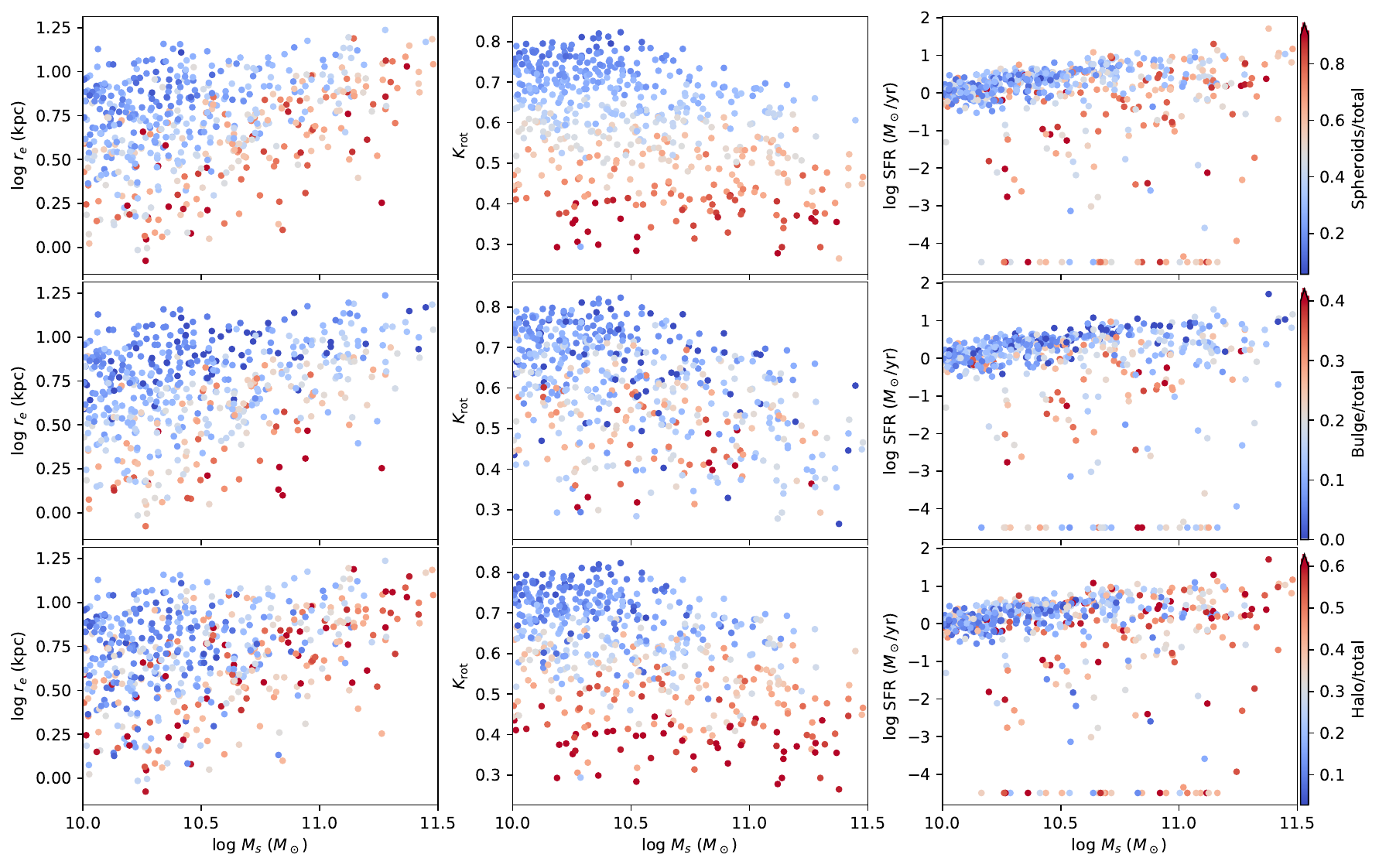}
\caption{Relation between global properties and the mass fractions of kinematically-derived structures for TNG50 central galaxies at $z=0$. From left to right, we show the distributions of the stellar half-mass radius $r_{\rm e}$, global rotation parameter $K_{\rm rot}$, and global instantaneous star formation rate as a function of total stellar mass. The same set of galaxies are coloured by the mass fractions of their spheroidal, bulge, and halo structures, respectively, from top to bottom. The cases of zero SFR are set to -4.5 in the left panels.}
\label{fig:TNG50basic}
\end{center}
\end{figure*}

\subsection{Global properties}
\label{sec:basic}

It is expected that intrinsic structures in galaxies are reflected by their morphological and kinematic 
properties, but possibly in a non-linear way. In this section, we discuss the relation between galaxies 
classified by our kinematic method and their global morphological and kinematic properties. 

Figure \ref{fig:TNG50basic} shows some basic properties: stellar 
half-mass radius $r_{\rm e}$, global rotation $K_{\rm rot}$
\citep{Sales2010}, star formation rate (SFR hereafter), and the mass fractions of 
kinematic structures (see \reffig{fig:TNG100basic} for TNG100 galaxies in appendix \ref{appendix}). 
$K_{\rm rot}=\langle v_\phi^2/v^2 \rangle$, where $v_\phi$ and $v$ are the cylindrical 
rotation velocity and total velocity, respectively, quantifies the relative importance of cylindrical 
rotation. Each data point is coloured by the mass fraction of its spheroid, bulge, and halo, 
respectively, from top to bottom. Clearly, galaxies with different kinematic structures have very different 
properties. This is a confirmation of the bounty of both the galaxy formation model underlying TNG50 as 
well as our kinematically-motivated stellar decomposition method. The galaxies that are dominated by 
spheroidal structures (red points in the top panels) generally have relatively compact morphologies, 
quiescent SF and weak rotation. The blue points, mainly corresponding to galaxies dominated by 
kinematic disks, are generally extended galaxies with active SF and strong rotation, that preferentially 
populate the low-mass end ($M_{\rm s} < 10^{10.6} M_\odot$) of the distribution. 

It is clearly shown in the top-left panel of \reffig{fig:TNG50basic} that galaxies dominated by spheroidal 
structures are common in massive galaxies producing the well-known mass-size relation, where disk 
galaxies are rare. Systematic comparisons between the mass-size relation in observations and that in \TNG\ 
are given in \citet{Rodriguez-Gomez2019, Genel2018} for TNG100 and in \citet{Pillepich2019} for TNG50. As 
shown in the middle panels of \reffig{fig:TNG50basic}, galaxies with more massive bulges are generally more 
compact (smaller size and larger central density), while those with massive halos (bottom panels) are not 
that dramatically different in $r_{\rm e}$ from galaxies dominated by disky structures. Interestingly, we 
can see that many galaxies with massive stellar halos are as extended as disk galaxies in less massive 
cases of $M_{\rm s} \lesssim 10^{10.6}$. Galaxies with massive bulges are generally the most compact 
objects over a broad mass range. 

The mass fraction of spheroidal components decomposed by \autoGMM\ is tightly correlated with 
$K_{\rm rot}$ that is almost independent of galaxy stellar mass. $K_{\rm rot}>0.5$ has been 
widely used as a criterion to select disk galaxies in simulations. This criterion selects almost the same 
group of galaxies as using $f_{\rm sph} < 0.5$. Both galaxies with massive bulges and halos are thus generally 
classified as elliptical/early-type galaxies, while they are clearly different types of galaxies. The galaxies 
with massive bulges have somewhat stronger rotation ($K_{\rm rot} \sim 0.4-0.6$) and more disky morphology 
(\reffig{fig:example_dens}) than those with massive halos. This suggests that galaxies with massive bulges are 
analogues of fast rotator ETGs from both morphological and kinematic points of view. But there is no clear 
dividing line in $K_{\rm rot}$ that can separate them from galaxies with massive halos that are slow rotator 
analogues. 

It is worth mentioning that, at the low-mass end, many galaxies dominated by spheroidal components are 
still actively forming stars, falling on the main sequence of disk galaxies (blue dots in the right panels of 
\reffig{fig:TNG50basic}). This result suggests the quenching is unlikely to be directly correlated with the 
growth of either bulges or halos in central galaxies. It is worth emphasizing that many star-forming galaxies 
with massive spheroids are Sombrero analogues in observations, but cannot be distinguished easily from spiral 
galaxies. Disks in such galaxies generally still have at least 20 per cent of their 
total stellar masses. This fraction may be even larger using the classical bulge-disk decomposition in 
morphology due to the contamination of stellar halos in low-inclination cases, as suggested 
by \citet{Du2020}. Nevertheless, spiral structures are commonly visible in the face-on view. 
Sombrero analogues are, thus, likely to be classified as disk galaxies, instead of elliptical ones. 
A systematic study is required to make a robust conclusion on whether SFRs are less 
sufficiently suppressed in elliptical galaxies from the \TNG\ simulations.

\begin{figure*}[htbp]
\begin{center}
\includegraphics[width=0.48\textwidth]{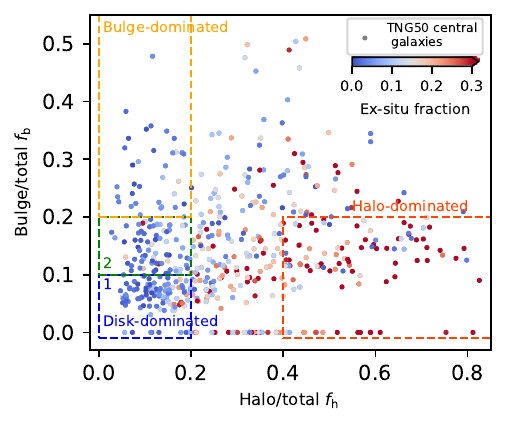}
\includegraphics[width=0.48\textwidth]{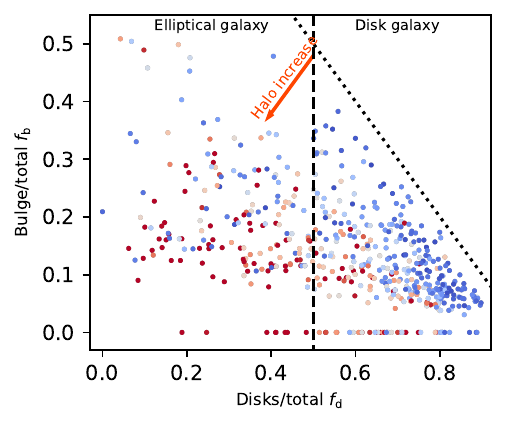}
\caption{Mass fractions of kinematic structures in TNG50 central galaxies and definition of disk-, bulge, and stellar halo-dominated galaxies. The color represents the mass fraction of stars formed ex situ. The rectangles in the left panel highlight the three groups of galaxies, we selected, that are dominated by disks, bulges, and halos. Disk-dominated galaxies are further divided into two sub-groups. Shown in the right panel, we see two branches with the decrease of mass fraction of disky components: those in the lower branch have been significantly affected by mergers, while mergers rarely happen for those in the upper one. The dashed line marks the criterion of disk and elliptical galaxies $f_{\rm sph} = 0.5$. 
It is worth mentioning that the cases of $f_{\rm b} = 0$ have no bulges that are prominent enough to be identified by the kinematic decomposition method. The gap around $f_{\rm b} \sim 0.03$ is thus not physically meaningful.}
\label{fig:classTNG50}
\end{center}
\end{figure*}

\begin{figure}[htbp]
\begin{center}
\includegraphics[width=0.48\textwidth]{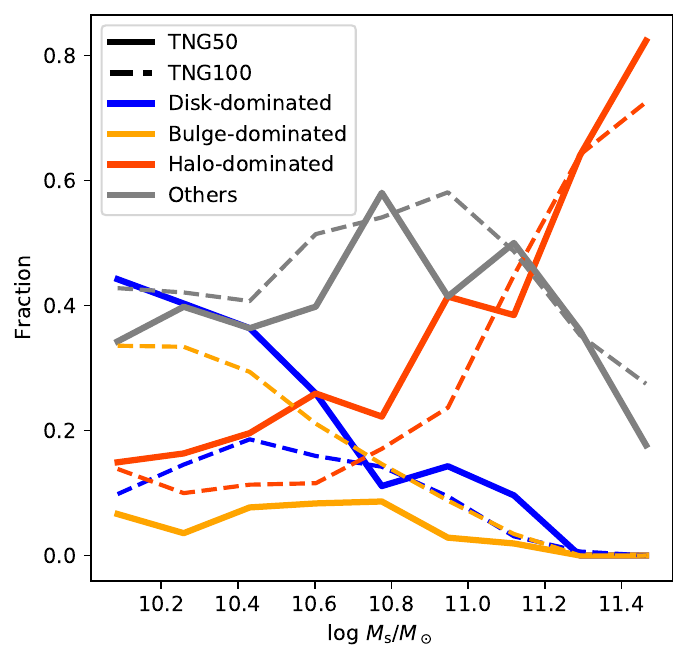}
\caption{Number fraction of galaxies in each group defined in Figure 7. Solid and dashed profiles show the results of TNG50 and TNG100, respectively.}
\label{fig:population}
\end{center}
\end{figure}

\begin{figure*}[htb]
\begin{center}
\includegraphics[width=0.99\textwidth]{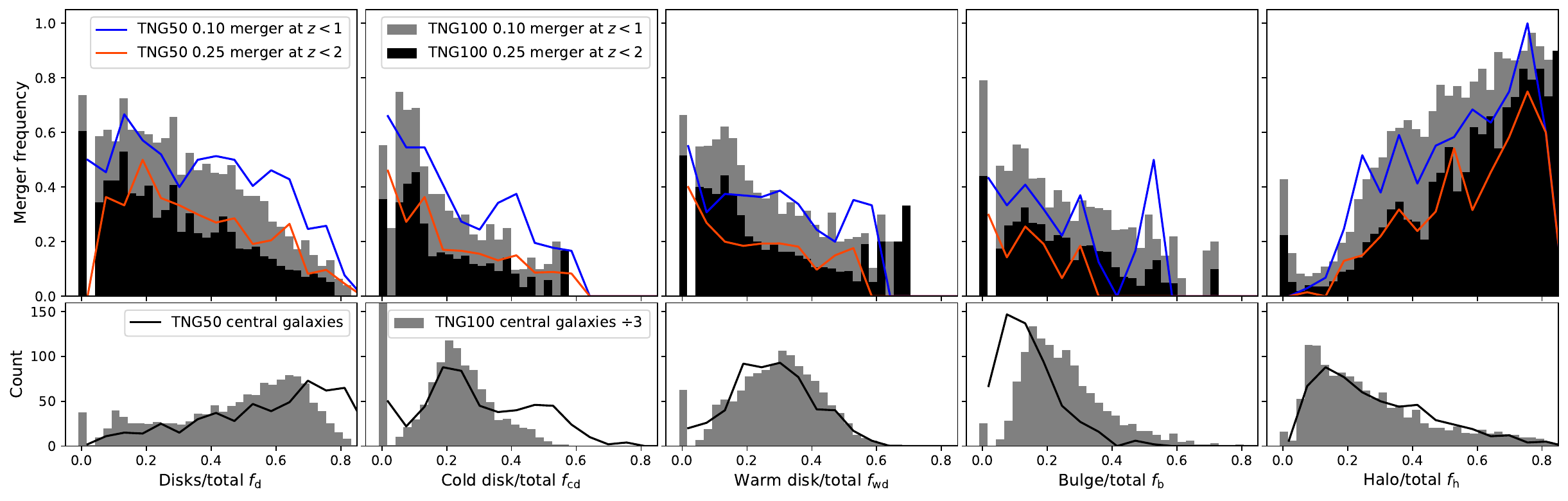}
\caption{Merger frequency (upper panels) in galaxies ($10^{10}\ M_\odot \leq M_{\rm s} \leq 10^{11.5}\ M_\odot$) with different kinematic structures. It measures the fraction of galaxies that have experienced at least one certain merger since a particular redshift. Here the mergers of mass ratio $\geq 0.1$ (1:10 minor mergers) at $z<1$ and $\geq 0.25$ (1:4 major mergers) at $z<2$ are taken into account. $f_{\rm d}$ is equal to $f_{\rm cd} + f_{\rm wd}$. The lower panels show the number counts of galaxies in each bin. The number distribution of TNG100 galaxies is divided by three to compare with that of the TNG50 galaxies.}
\label{fig:centh}
\end{center}
\end{figure*}
\subsection{Connection to merger history}
\label{subsec:merger}

\section{A kinematic selection of galaxies dominated by disks, bulges, and halos}
\label{sec:kinemclass}

\subsection{Definition of \\disk-, bulge-, and halo- dominated galaxies}

\reffig{fig:classTNG50} shows the mass fractions of kinematic structures for all central galaxies from 
TNG50 in the $10^{10-11.5}\ M_\odot$ stellar mass range. The ratios of stellar halo, bulge, and disk 
mass, respectively, to the total stellar mass are denoted with $f_{\rm h}$, $f_{\rm b}$, and $f_{\rm d}$.
In the left panel of \reffig{fig:classTNG50}, we select three groups of galaxies:
\begin{itemize}
\item {\it Disk-dominated 1 and 2:} $f_{\rm h} < 0.2$. For the groups 1 (blue rectangle) and 2 (green rectangle), $f_{\rm b}$ is $< 0.1$ and $0.1-0.2$, respectively.  
\item {\it Bulge-dominated:} $f_{\rm b} \geq 0.2$ and $f_{\rm h} < 0.2$, orange rectangle.
\item {\it Halo-dominated:} $f_{\rm b} < 0.2$ and $f_{\rm h} \geq 0.4$, red rectangle. 
\end{itemize} 

From left to right in the right panel, the mass fraction of disky structures increases, thus galaxies 
change from spheroidal early-type to disky late-type ones. A large group of galaxies dominated by disks 
clusters at the bottom-right corner in the right panel of \reffig{fig:classTNG50} (also the bottom-left 
corner in the left panel). These galaxies are akin to pure-disk/bulgeless galaxies in observations (see 
their edge-on view in \reffig{fig:example_dens}). Note that the mass fraction of disks $f_{\rm d}$ is 
obtained by summing all stars in their cold and warm disks (which includes any disky/pseudo bulge). 
The mass fraction decrease of disky structures leads to the increase of either a bulge or a halo, 
thus two branches. Galaxies on the lower branch have relatively more massive halos. 

\reffig{fig:population} shows the fraction of each galaxy group as a function of stellar mass. It is 
clear that the relatively low-mass galaxies (stellar mass $<10^{10.6}M_\odot$) in TNG100 have much more 
bulge-dominated cases. Star formation is generally still active in such galaxies. It is thus consistent 
with the conclusion of \citet{Rodriguez-Gomez2019} that TNG100 produces many blue spheroids. Similarly,
\citet{Du2020} showed that disk galaxies in TNG100 generate a dramatically lower fraction of cold disks 
than those extracted in CALIFA galaxies \citep{Zhu2018b}. The result in this paper suggests that this issue 
has been significantly improved in TNG50, if not completely solved, possibly due to its better resolution.

In total, among the 541 central TNG50 galaxies, 183 ($\sim 34\%$) are disk-dominated. This fraction drops to 
14\% (75 galaxies) if we only count the galaxies that are dominated by cold disks (mass fraction 
$f_{\rm cd}\geq0.5$), which is a conservative definition of disk-dominated galaxies. About 1\% 
(5 galaxies) have cold disks with $f_{\rm cd}\geq0.7$.
The conclusions in \refsec{sec:bulgeanddisk} will not be significantly affected if we 
distinguish bulge- and disk-dominated galaxies according to the mass fraction of their cold disks.

Dissipationless ``dry'' mergers in the later phase are expected to be destructive for disky structures. 
In \reffig{fig:classTNG50}, individual simulated galaxies are color-coded by the amount of stellar mass 
that was not formed in the galaxies lying along the main progenitor branch in the merger trees, i.e., ex-situ, 
estimated using the method of \citet{Rodriguez-Gomez2016}. It is clear that the mass fraction of ex-situ 
stars is generally $<0.1$ in both disk- and bulge-dominated galaxies, while it is much larger in halo-dominated galaxies. 

In the upper panels of \reffig{fig:centh}, we show the merger frequency of galaxies with different kinematic structures. 
It measures the fraction of galaxies that have experienced at least one merger of certain mass ratio since a 
particular redshift. The lower panels show number counts of galaxies. Clearly, the mass fraction of a kinematic stellar halo has a 
strong positive correlation with mergers. About 80 per cent of the central galaxies with massive stellar halos of 
mass fraction $f_{\rm h} > 0.4$ have experienced at least one merger of stellar mass ratio $\geq 0.1$ since $z=1$ 
(see blue curves for TNG50 and gray histograms for TNG100). More than 50 per cent of such galaxies are associated 
with 0.25 major mergers (red profiles for TNG50 and black histogram for TNG100), in agreement with 
\citet{Penoyre2017} and \citet{Lagos2018}.

Despite the somewhat artificial classification, both disk- and bulge-dominated galaxies have rarely been affected 
by mergers in the past 10 Gyr ($z<2$), thus they are likely to be two fundamentally-different types of galaxies in 
comparison to halo-dominated ones. Mergers have been infrequent in the late-phase of cosmic evolution for 
bulge-dominated galaxies, so that such bulges must have been generated by either internal processes or in an earlier-phase 
of the Universe. Bulge- and disk-dominated galaxies may be the two ends of a continuous distribution. The properties 
of such galaxies are able to record important information of their ``initial conditions'' after the early phase, 
which is likely lost in mergers. The formation of halo-dominated galaxies, in contrast, is tightly correlated with 
mergers that happen late, thus are relatively dry. They may form via major mergers between the two fundamental 
types of galaxies, or via ex-situ stellar  accumulation through minor mergers with low-mass satellites. In order to 
gain new insights into galaxy properties discussed here and their evolutionary histories, we trace different types 
of galaxies back to high redshifts in the next two sections.

\section{Evolution of disk- and bulge-dominated galaxies driven by the early-phase and internal processes}
\label{sec:bulgeanddisk}

Given the mass fractions of kinematically-derived structures, we are able to study the difference in 
their evolutionary histories in detail for galaxies with different structures. The most fundamental 
physical processes are separated into three parts: the
early phase evolution, and internal and external processes in the later phase. 
The rich diversity in kinematic structure will be interpreted in the context of these three origins. 
In this work, we regard all processes at $z>2$ as the early-phase evolution of galaxies. At $z<2$, 
galaxy evolution can be influenced by internal and external (mainly but not exclusively mergers) 
processes. 
The {\it SubLink} galaxy merger tree of the \TNG\ simulations 
\citep{Rodriguez-Gomez2015} is used to trace galaxy evolution back in time. 

\begin{figure*}[htbp]
\begin{center}
\includegraphics[width=0.99\textwidth]{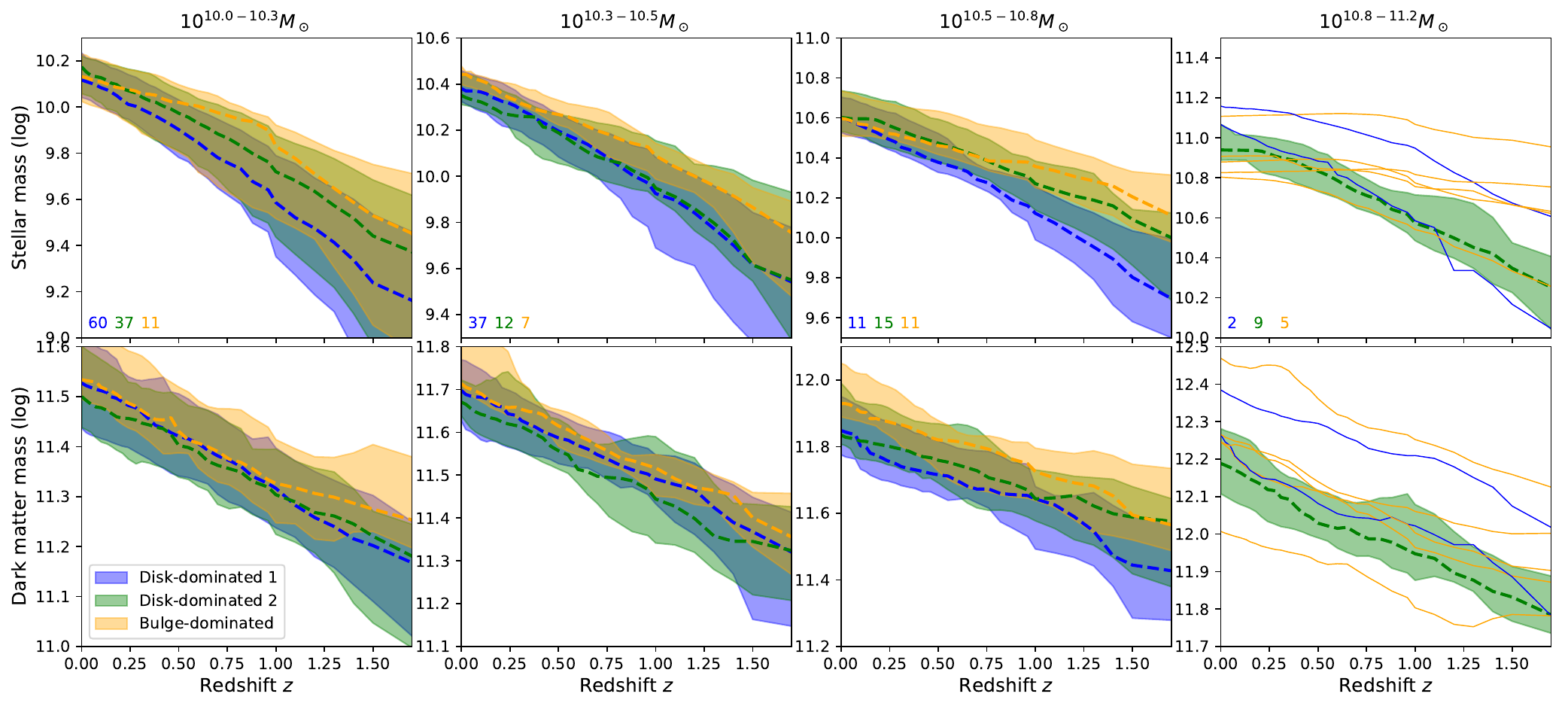}
\caption{Mass growth ($y$-axis, in logarithmic scale) of stellar and dark matter components in TNG50 central galaxies classified by the kinematic method. From left to right, galaxies are separated into four mass bins using the total stellar mass at $z=0$. The shaded regions represent the stacked 1D profiles ($1\sigma$ envelope) for the disk-dominated (group 1 and 2) and bulge-dominated galaxies, in each mass bin, where the dashed profiles correspond to their median values. From left to right, galaxies are shown in four mass bins of their stars at $z=0$, i.e., $10^{10-10.3}\ M_\odot$, $10^{10.3-10.5}\ M_\odot$, $10^{10.5-10.8}\ M_\odot$, and $10^{10.8-11.2}\ M_\odot$, respectively. The solid profiles in the right-most panels show each individual galaxy when the statistics are poor in that mass bin. The number of galaxies is listed at the bottom-left corner.} 
\label{fig:growth}
\end{center}
\end{figure*}

\begin{figure*}[htbp]
\begin{center}
\includegraphics[width=1.0\textwidth]{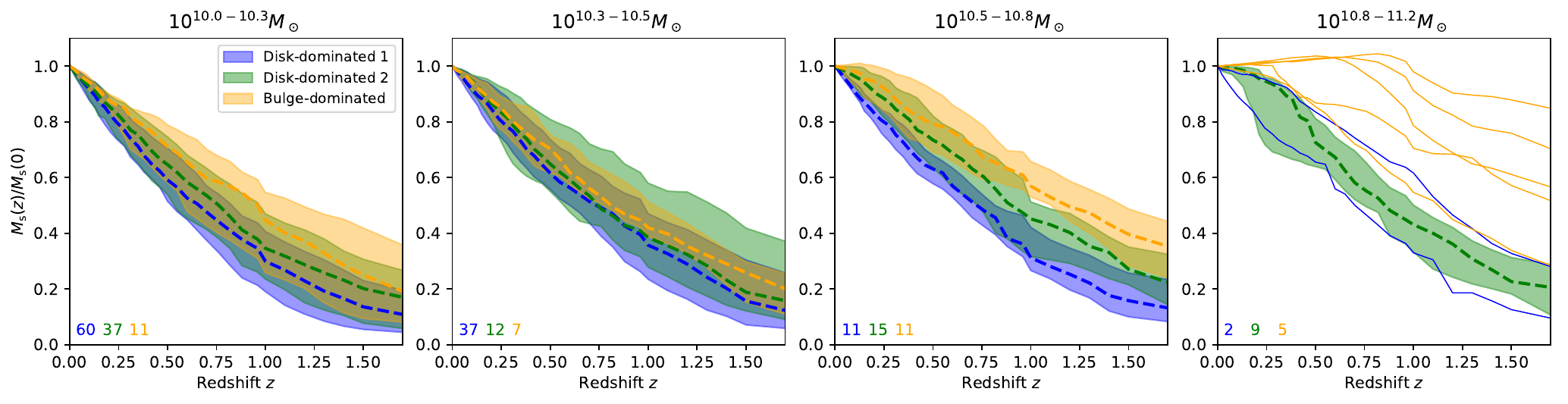}
\caption{Evolution of total stellar mass $M_{\rm s}$, normalized by the values at $z=0$. This image uses the same convention as \reffig{fig:growth}.}
\label{fig:growthratio}
\end{center}
\end{figure*}

\begin{figure*}[htbp]
\begin{center}
\includegraphics[width=\textwidth]{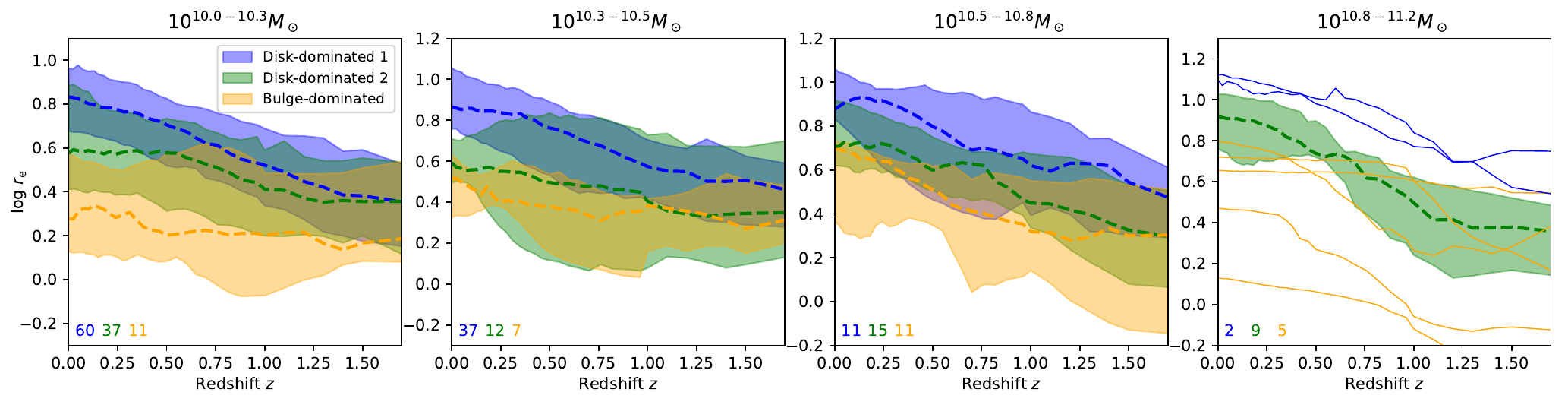}
\caption{Evolution of half-mass radius $r_{\rm e}$, measured in the three-dimensional space. This image uses the same convention as \reffig{fig:growth}.}
\label{fig:size}

\includegraphics[width=\textwidth]{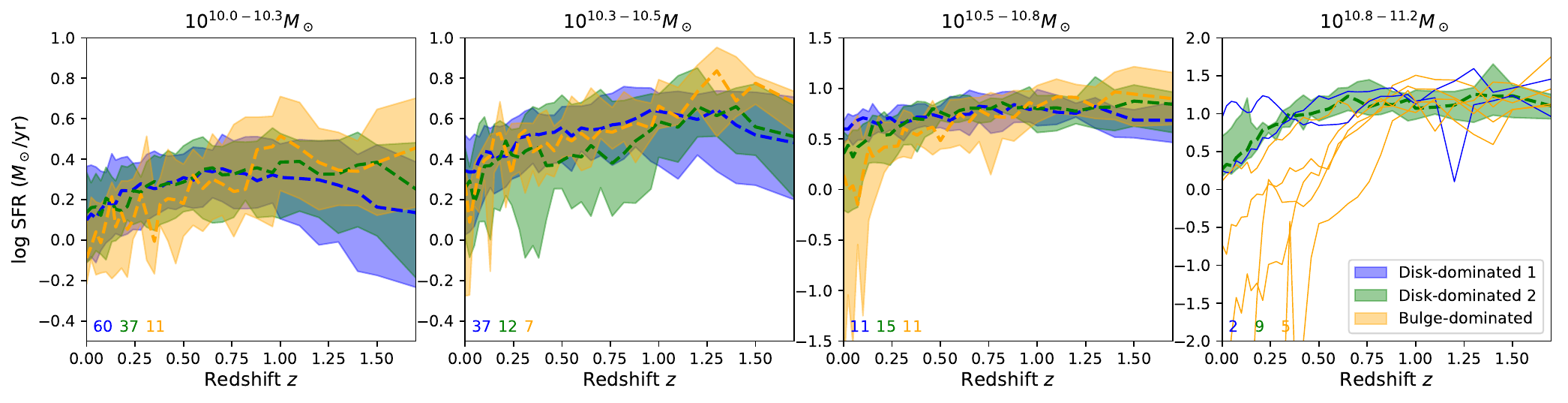}
\caption{Evolution of total SFR. This image uses the same convention as \reffig{fig:growth}.}
\label{fig:SFRevo}

\includegraphics[width=\textwidth]{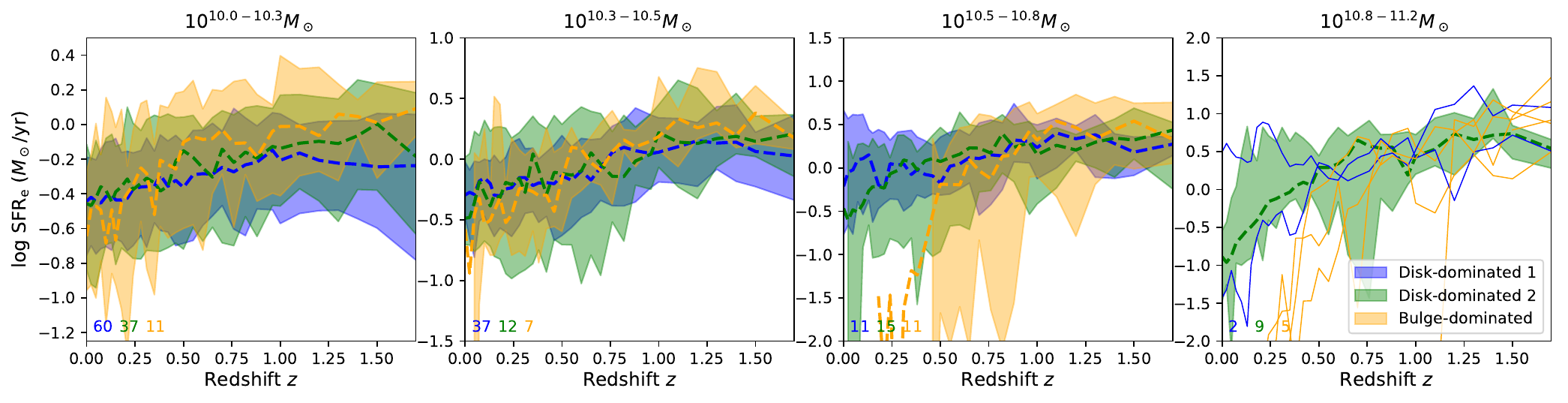}
\caption{Evolution of the SFR within one $r_{\rm e}$. This image uses the same convention as \reffig{fig:growth}. Massive bulge-dominated galaxies 
are quenched in an inside-out manner in comparison with \reffig{fig:SFRevo}.}
\label{fig:SFRre}

\includegraphics[width=\textwidth]{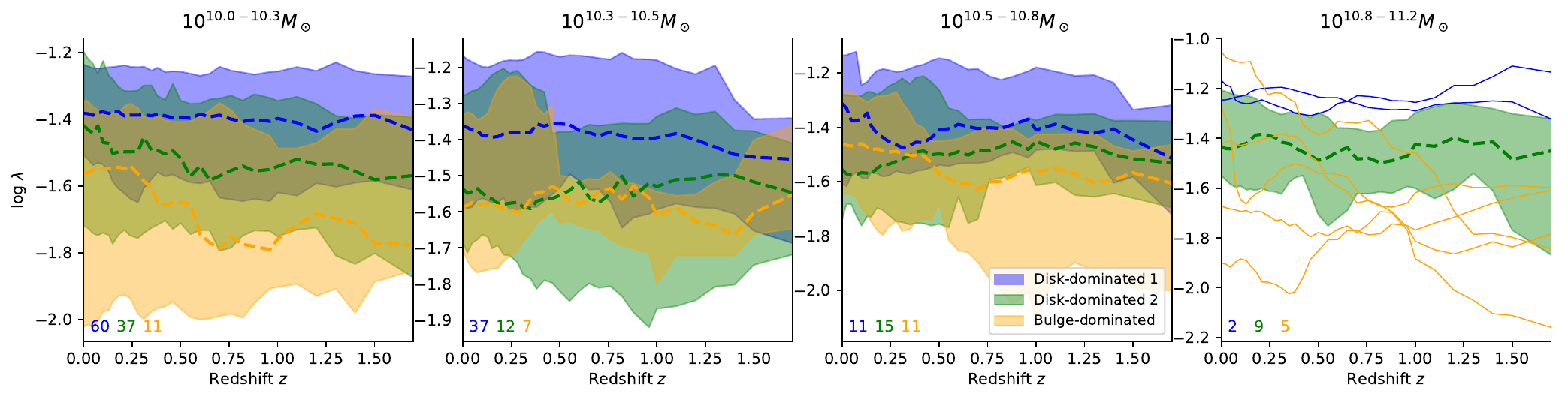}
\caption{Evolution of the dimensionless spin parameter $\lambda$, derived all member particles/cells, including star, dark matter, and gas. This image uses the same convention as \reffig{fig:growth}.}
\label{fig:spin}
\end{center}
\end{figure*}

\begin{figure}[htbp]
\begin{center}
\includegraphics[width=0.5\textwidth]{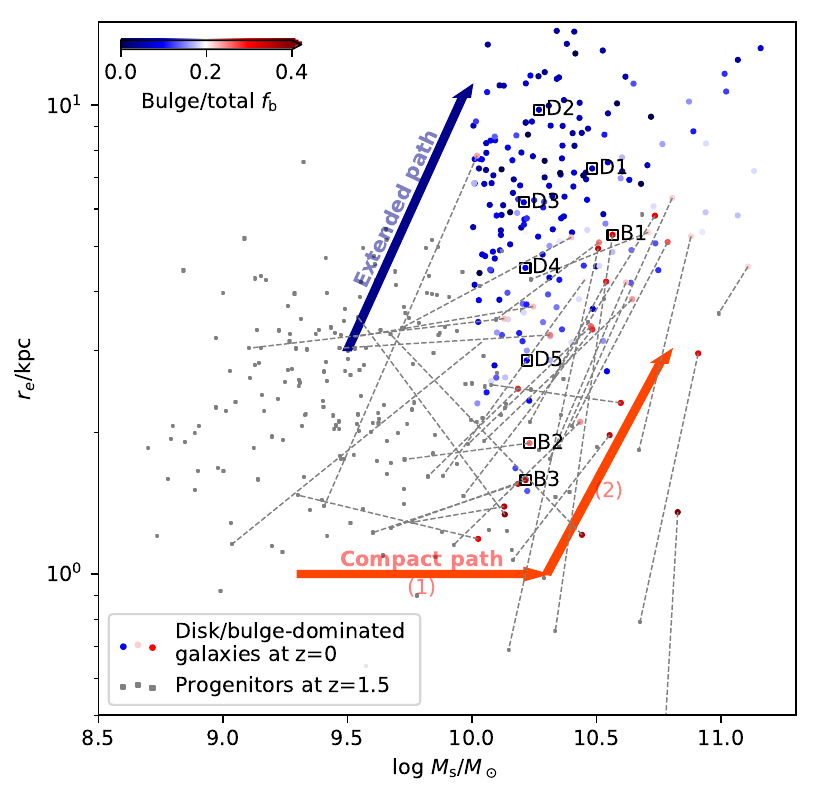}
\caption{The mass-size diagram of the disk- and bulge-dominated galaxies that 
are the two fundamental types of galaxies selected in \reffig{fig:classTNG50}. The color 
represents the bulge-to-total mass fraction $f_{\rm b}$ derived by our kinematic method. The 
gray dots are their progenitors at $z=1.5$. Dashed lines mark the evolutionary pathway during $z=0-1.5$ 
for all bulge-dominated galaxies. The arrows highlight the extended (blue) and compact (red) 
evolutionary pathways that form disk and bulge-dominated galaxies, respectively. 
Prototype galaxies shown in Figures \ref{fig:D1}, \ref{fig:B1}, \ref{fig:gas}, and \ref{fig:B2} are 
marked by squares.} 
\label{fig:DBpath}
\end{center}
\end{figure}

\subsection{Extended and compact evolutionary pathways: Evolution of mass, size, SFR, and spin}

\subsubsection{Mass and size}
In \reffig{fig:growth}, we trace the mass growth of the stellar and dark matter components in each galaxy. 
The evolution of disk- and bulge-dominated galaxies generally follow 
smooth evolutionary pathways without experiencing any violent mergers, as shown in \reffig{fig:classTNG50}. 
In each mass range, we thus stack their profiles together. Galaxies dominated by disks (blue and green shaded regions) 
form later, but grow faster, compared with those dominated by bulges (cyan shaded regions), thus 
reaching a similar stellar mass at $z=0$. At $z>1.0$, the difference in median stellar mass between 
the group 1 disk-dominated galaxies and bulge-dominated galaxies is about 0.2-0.4 dex over a wide mass 
range; this difference decreases 
gradually toward low redshifts. The difference is more significant in more massive galaxies 
(e.g., $M_{\rm s}=10^{10.5-10.8}\ M_\odot$), shown also clearly in \reffig{fig:growthratio} where 
we normalize the stellar masses of galaxy progenitors with the value at $z=0$. About 30-55\% stellar mass 
has been assembled at $z\sim 1.7$ in massive bulge-dominated galaxies with $M_{\rm s}=10^{10.5-10.8}\ M_\odot$, 
while only about 5-25\% of stars exist in the progenitors of the group 1 disk-dominated galaxies. 

Galaxy size is another crucial parameter that reflects various physical processes in the evolutionary 
history of galaxies. \citet{Pillepich2019} showed that TNG50 successfully reproduces the mass-size 
relation with respect to the observations of both gaseous \citep[e.g.][]{vanderWel2014} and stellar 
components across cosmic time. 
\reffig{fig:size} exhibits the growth of galaxy size. At $z=0$, the disk-dominated galaxies in group 1 have 
roughly 2-3 times larger stellar half-mass radius $r_{\rm e}$ than the bulge-dominated galaxies. At high 
redshifts, the difference in their sizes is smaller, but bulge-dominated galaxies are generally a few times 
more massive than disk-dominated ones. Thus, bulge-dominated galaxies are much more compact objects 
than disk-dominated galaxies. Such compact and extended types of galaxies follow very different 
evolutionary pathways, then generate bulge- and disk-dominated galaxies, respectively.
Consistently, \citet{Genel2018} showed that the sizes of star-forming and quiescent galaxies from TNG100 
evolve in similar extended and compact pathways. 

\subsubsection{Star formation}
In the later phase, bulge-dominated galaxies start to be quenched gradually, especially more massive systems.
During $z\lesssim0.5$, most massive ($M_{\rm s}\gtrsim10^{10.5}\ M_\odot$) bulge-dominated galaxies increase by less 
than 30\% of their total stellar mass at $z=0$, while in disk-dominated galaxies stellar masses are nearly doubled 
(see \reffig{fig:growthratio}). \reffig{fig:SFRevo} shows clearly that massive bulge-dominated galaxies start 
to be quenched significantly since $z\lesssim 1.0$, thus offset from their disk-dominated counterparts. 
Quenching happens later and is less significant in less massive galaxies. Massive bulge-dominated 
galaxies are likely quenched by AGN feedback, as a consequence of activating the low accretion 
(kinetic) mode of AGN feedback \citep{Weinberger2017, Weinberger2018, Nelson2019, Terrazas2020}. 
In this case, mass outflow rates 
increase rapidly, which pushes gas out and then quenches SF gradually. This mechanism is insufficient 
in quenching less massive bulge-dominated galaxies (the left-most panels of \reffig{fig:growth}), where 
the black hole mass is generally smaller. Moreover, in comparison with the SFR measured within 
$r_{\rm e}$ (\reffig{fig:SFRre}), it is clear that SF is quenched in an inside-out manner (see also 
\citet{Nelson2019} and \citet{Nelson2021}) in massive bulge-dominated galaxies. 

Disk- and bulge-dominated galaxies follow two distinguishable evolutionary pathways: extended and compact. 
A massive bulge forms either earlier or more easily in bulge-dominated galaxies, thus are more compact than 
disk-dominated ones. Such a difference can only be interpreted by the natural properties that are largely 
determined by the dark matter halos they inhabit and underlying internal dynamical instabilities 
(see more discussions in \refsec{sec:nature}). 

\subsubsection{Spin}
The mass and angular momenta of dark matter halos are two crucial factors that may significantly affect 
galaxy properties. Here we characterize the angular momentum of a halo with the dimensionless spin parameter 
$\lambda=\frac{j}{\sqrt{2}V_{\rm vir}R_{\rm vir}}$ \citep{Bullock2001} where $V_{\rm vir}$ and $R_{\rm vir}$ 
are virial velocity and radius estimated by the total mass bound to the halo. $j$ is the specific angular 
momentum of all member particles/cells, including stars, dark matter, and gas. We have confirmed that such 
a spin parameter has a small difference from the one derived for only dark matter particles that have 
not been saved for every snapshot.
It is clear that bulge-dominated galaxies are generally present in systems with significantly 
lower spins. \citep[defined by][see \reffig{fig:spin}]{Bullock2001}, though the
dark matter halos (lower panels of \reffig{fig:growth}) that bulge-dominated galaxies inhabit are also 
somewhat more massive. This is consistent with the straightforward theoretical picture that gas 
initially coupled with the dark matter haloes cools down to the center while conserving angular 
momentum \citep{Mo1998, Bullock2001}, possibly by a certain factor. It finally leads to the bimodality in galaxy compactness 
\citep[see][]{Dekel&Burkert2014} that assembles together via the dissipative processes in the early phase. 
However, whether the angular momentum can be conserved sufficiently is still under debate. \citet{Jiang2019} did not 
find a clear correlation for galaxies from zoom-in hydro-cosmological simulations, which may be due to 
the loss of angular momentum when cold streams falling into the inner regions of halos \citep{Danovich2015}. 
Bulge- and disk-dominated galaxies in TNG100 galaxies follow a similar evolution to those in 
TNG50, shown in the appendix \ref{appendix} (\reffig{fig:TNG100evo}).

\begin{figure*}[htbp]
\begin{center}
\includegraphics[width=0.95\textwidth]{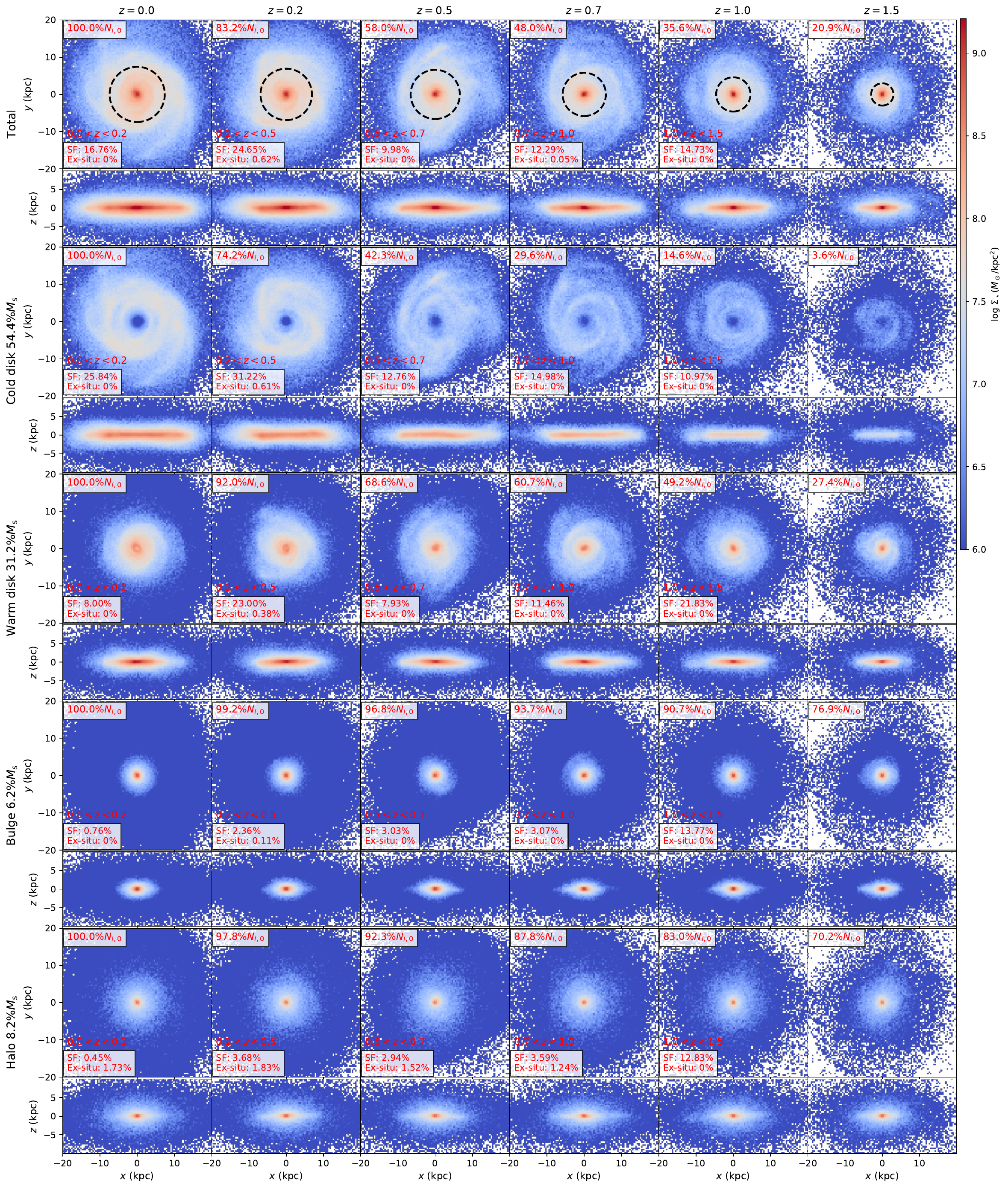} 
\caption{D1, a disk-dominated galaxy (ID 580035) with $M_{\rm s}\approx 10^{10.5} M_\odot$. From top to bottom, we show the evolution of total, cold disk, warm disk, bulge, and halo stellar structures, decomposed by our kinematic method, in both face-on and edge-on views. Their mass fractions are given on the left side. The 3D half-mass radius $r_{\rm e}$ at each snapshot is marked by the dashed circle. In each face-on panel, the top textbox gives the fraction of stellar particles that already exist in this galaxy at this snapshot for each structure. In the bottom textbox, we estimate the contributions of in-situ SF and ex-situ mergers to the mass growth of each structure in a time span between two snapshots. In this galaxy, the kinematic halo and bulge are small. In-situ SF overall dominates the evolution of all structures. The contribution from ex-situ processes is  negligible since $z=1.5$.}
\label{fig:D1}
\end{center}
\end{figure*}

\begin{figure*}[htbp]
\begin{center}
\includegraphics[width=0.95\textwidth]{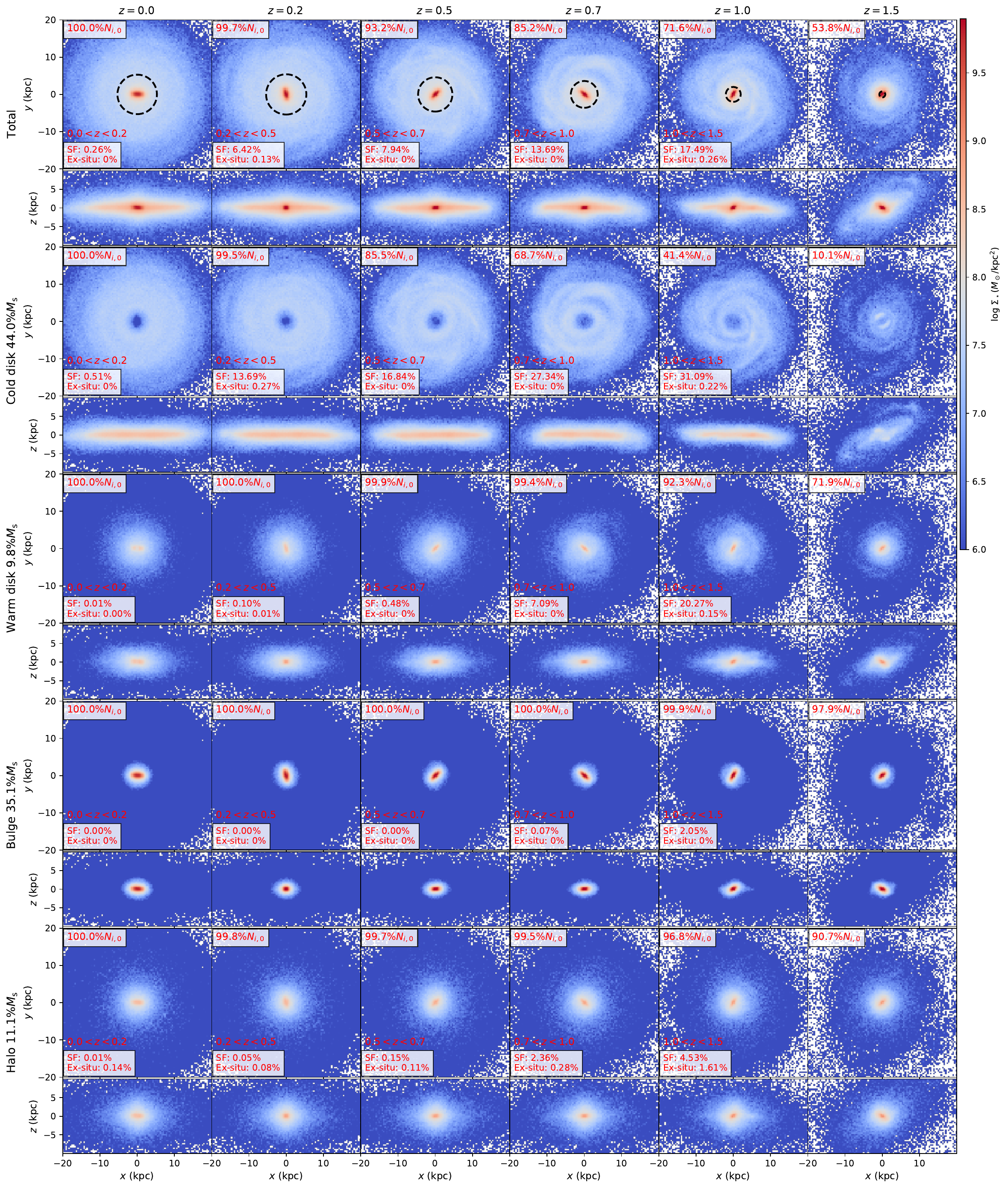} 
\caption{B1, a bulge-dominated galaxy (ID 563732) with $M_{\rm s}\approx 10^{10.5} M_\odot$.This image uses the same convention as \reffig{fig:D1}. The kinematic bulge forms early with no influence from mergers since $z=1.5$, then a disk assembles, possibly through gas accretion in the later phase. The mis-aligned disk at $z\sim 1.5$ is likely due to mis-aligned cold gas accretion or tiny mergers. This galaxy can be a S0 galaxy with a compact classical-like bulge by visual classification. }
\label{fig:B1}
\end{center}
\end{figure*}

\begin{figure*}[htbp]
\begin{center}
\includegraphics[width=0.95\textwidth]{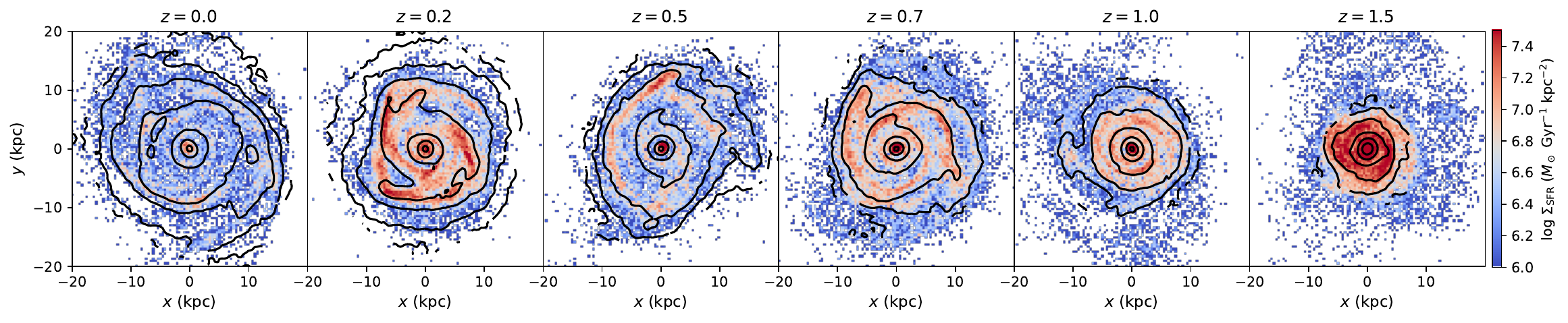} 
\includegraphics[width=0.95\textwidth]{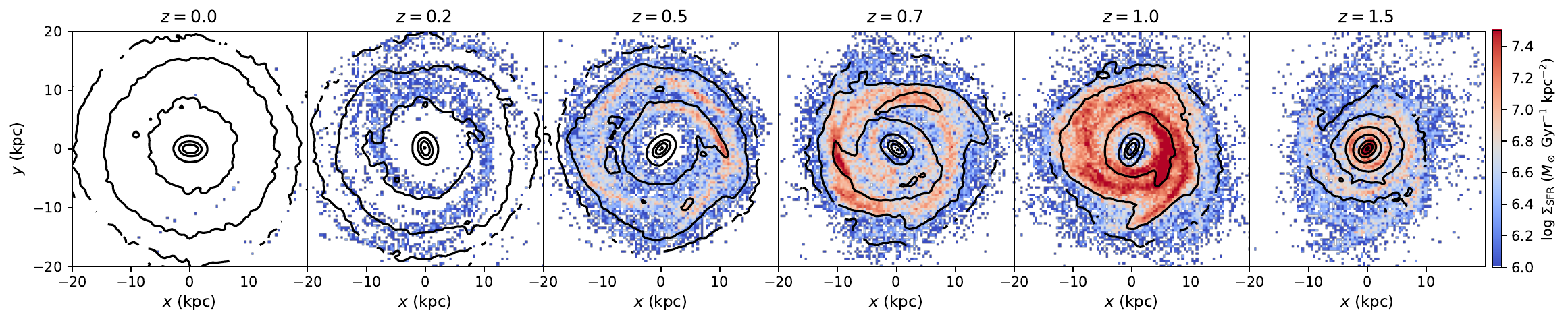} 
\caption{Spatial distributions of SFRs in D1 (top) and B1 (bottom), averaged within 0.5 Gyr. The black contours represent the stellar surface density maps. Clearly, B1 is quenched inside-out. The size increases of D1 and B1 are mainly driven by the assembly of their disky structures via SF in the outer regions.}
\label{fig:SFR_D1B1}
\end{center}
\end{figure*}

\subsection{Disk growth in massive cases: gas accretion and inside-out quenching}

\reffig{fig:DBpath} shows the extended and compact pathways on the mass-size diagram. The color 
represents the mass fraction of bulges in both disk- and bulge-dominated galaxies at $z=0$; the 
gray data points correspond to their progenitors at $z=1.5$. Disk-dominated galaxies generally follow the 
extended pathway, highlighted by the blue arrow. Bulge-dominated galaxies follow the compact pathway 
that has two phases: (1) a compact phase during which the mass grows significantly 
while the size changes little, thus forming bulges; (2) the size increases significantly while 
the mass grows relatively little, thus building up disky structures. 
Massive cases with $M_{\rm s} \gtrsim 10^{10.3}\ M_\odot$ have almost passed through the phase (1) 
at $z\sim 1.0$, while the compact phase seems to last to $z=0$ for less massive cases. 

The progenitors of massive bulge-dominated galaxies (linked by the dashed-gray lines) 
with $M_{\rm s}\gtrsim 10^{10.5}\ M_\odot$ have already been rather massive and compact at $z=1.5$. 
They then evolve into the phase (2) of the compact pathway, during which diffuse disk structures are 
assembled gradually, thus their sizes increase in a similar way to disk-dominated ones. Figures 
\ref{fig:D1} and \ref{fig:B1} show two prototypes of massive disk- and bulge-dominated galaxies, 
named D1 and B1 (marked by squares in \reffig{fig:DBpath}), respectively, to illustrate 
the dramatically different evolutionary pathways between them. Stellar particles are classified into 
the structure that has the largest likelihood at $z=0$ via applying our kinematic decomposition 
algorithm. From top to bottom, we show the surface density maps in both face-on and edge-on 
views for total, cold disk, warm disk, bulge, and halo.

Three quantities (red words in the top and bottom textboxes of Figures \ref{fig:D1} and \ref{fig:B1})  
are used to characterize the number fractions of stars that originate from an earlier-phase evolution, 
external/ex-situ mergers, and internal/in-situ SF (from top to bottom), respectively, for each kinematic 
structure (see details of their definitions in the footnote\footnote{$N_{i, z}$ is the total stellar particle 
number of a certain structure $i$ at redshift $z$. Tracing back from $z_1$ to an earlier time point $z_2$, the new 
stellar particle members of each structure during $z_1-z_2$ are classified into two origins: 
ex-situ accretion and SF, i.e., in-situ. The ex-situ part is estimated by the stars that 
are not belong to this galaxy at $z_2$; the in-situ part is newly formed stars in a time span 
between two snapshots from $z_2$ to $z_1$.}).
For example, the kinematic cold disk of D1 contributes $54.4\%$ of its total stellar mass at $z=0$. At $z=1.5$, 
only $3.5\%$ of cold disk stars, i.e., $3.5\%N_{i, 0}$, have already existed in this galaxy, 
where $N_{i, 0}$ is the total stellar particle number of a certain structure $i$ that is cold disk here. 
During $z=1.5-1.0$, it increases by $\approx 11.0\% N_{i, 0}$, where SF and ex-situ accretion contribute 
$10.97\% N_{i, 0}$ and $0\% N_{i, 0}$, respectively.  

Clearly, the properties at $z=1.5$ are largely dominated by their early-phase evolution. At $z=1.5$, 
the B1 object (\reffig{fig:B1}) has already assembled 53.8 percent of its stars found at $z=0$, while D1 
(\reffig{fig:D1}) has only had $20.9\%$ of its stars. The half-mass radius of B1, 
marked by dashed circles, is dramatically smaller than that of D1. A massive central 
concentration, i.e., bulge, is clearly visible in B1. The overall properties of the kinematic 
bulge in B1 changes mildly since $z=1.5$. Without experiencing mergers, the halo masses of both 
D1 and B1 also change little. 

During $z=0-1.5$, an extended cold disk forms gradually in both D1 and B1, which leads to the increase of 
their galaxy sizes. The growth of cold disks coincides well with the SF shown in \reffig{fig:SFR_D1B1}. 
At $z<1$, B1 is gradually quenched inside-out (see Figures \ref{fig:SFRevo} and \ref{fig:SFRre} for 
statistical results). An extended SF ring that is also gas-rich is formed. \citet{Zolotov2015} suggested 
that such a ring is a natural result of the accretion of cold gas with high angular momentum from the 
cosmic web into this node. Moreover, \citet{Dekel2020} showed that the existence of a massive 
central concentration, i.e., bulge, can suppress inward gas transport, which possibly also gives rise to
a SF ring. In conclusion, the existence of a massive bulge formed in the early phase evolution makes 
bulge-dominated galaxies more compact than disk-dominated ones. Both of them are able to 
generate similar disky structures by internal SF in the later phase. Massive bulge-dominated galaxies, 
however, are likely to be quenched at low redshifts, thus classified as fast-rotator ETGs with 
massive bulges.

\begin{figure}[htbp]
\begin{center}
\includegraphics[width=0.45\textwidth]{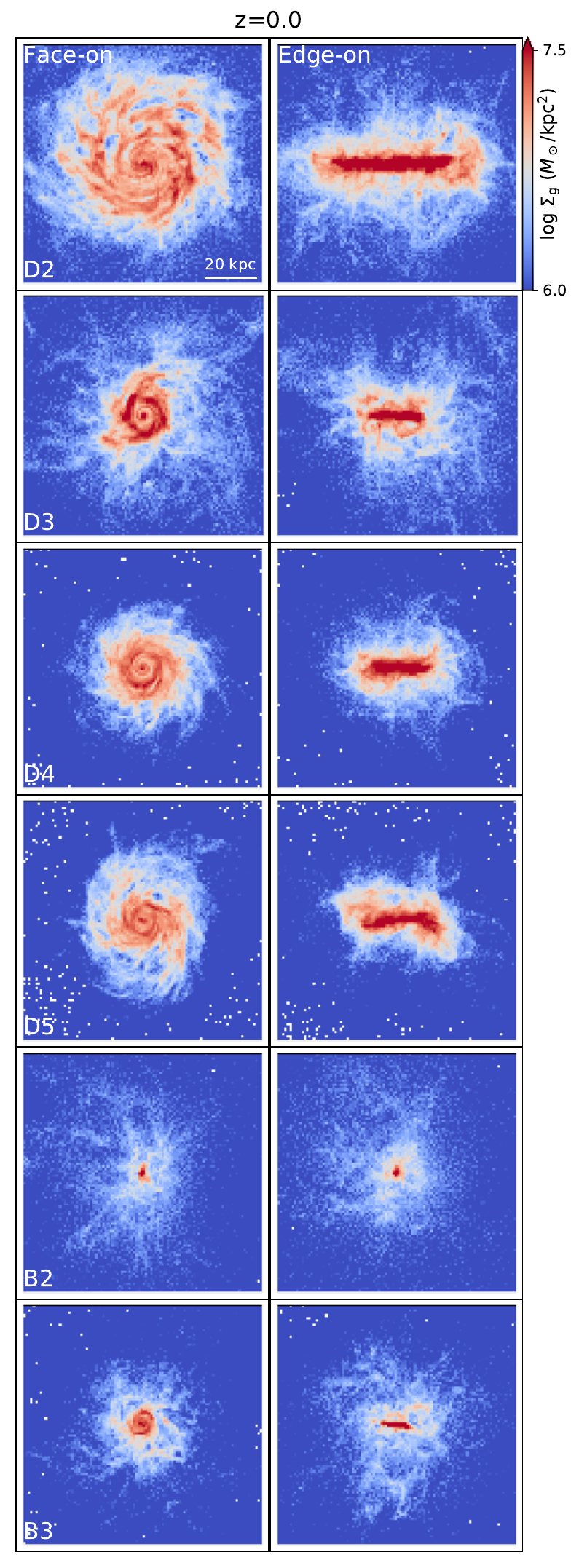}
\caption{Gas distributions in prototype galaxies D2-D4 and B2-B3, viewed face-on and edge-on. From D2 (top) to B3 (bottom), galaxies become more and more compact, which is likely determined by the gas that they accreted.} 
\label{fig:gas}
\end{center}
\end{figure}

\subsection{Size growth in less massive cases: undergoing bulge formation via gas accretion}
\label{sec:gasacc}
There is a dramatic difference in the size growth of disk- and bulge-dominated galaxies with 
$M_{\rm s} \lesssim 10^{10.3}\ M_\odot$, shown in Figures \ref{fig:SFRevo} and \ref{fig:DBpath}. 
The sizes of bulge-dominated galaxies are nearly flat or even decrease slightly while their stellar 
masses grow fast at $z \gtrsim 1$, while the $r_{\rm e}$ of disk-dominated galaxies increases significantly 
by a factor of $\sim 2.5$. However, it is surprising that the overall SFRs of such bulge- and 
disk-dominated galaxies follow a similar trend, though bulge-dominated galaxies 
have slightly higher SFR at $z>1$, but lower at $z<1$, than disk-dominated 
ones. The difference in SFR is not as significant as that in size between disk- and bulge-dominated 
galaxies. Therefore, the difference of compact and extended pathways must be due to the spatial 
distribution of their SF. Considering that both bulge- and disk-dominated galaxies are weakly 
affected by mergers, we speculate that their SF may be correlated with the cold gas inflows that are determined 
by the dark matter halos and environments.

\reffig{fig:gas} shows the face-on and edge-on views of gaseous mass in the disk-dominated galaxies 
D2-D5 and bulge-dominated galaxies B2-B3 (marked by squares in \reffig{fig:DBpath}). They have similar 
stellar masses of 
$M_{\rm s} \simeq 10^{10.2}\ M_\odot$, but their sizes vary from $\sim 1.5$ kpc to nearly 10 kpc. There is 
a clear signature that the dramatic difference in their sizes originates from the difference in angular 
momentum of accreted gas. More compact galaxies are likely to be fuelled by gas inflows 
whose angular momentum is lower or removed sufficiently, thus generating a smaller gaseous disk 
(\reffig{fig:spin}), which is consistent with the theoretical expectation \citep[e.g.][]{Mo1998, Bullock2001}.
Moreover, the galactic wind driven by supernova and AGN feedback may also play a role, as suggested 
by \citet{Genel2015}. 

\begin{figure*}[htbp]
\begin{center}
\includegraphics[width=0.95\textwidth]{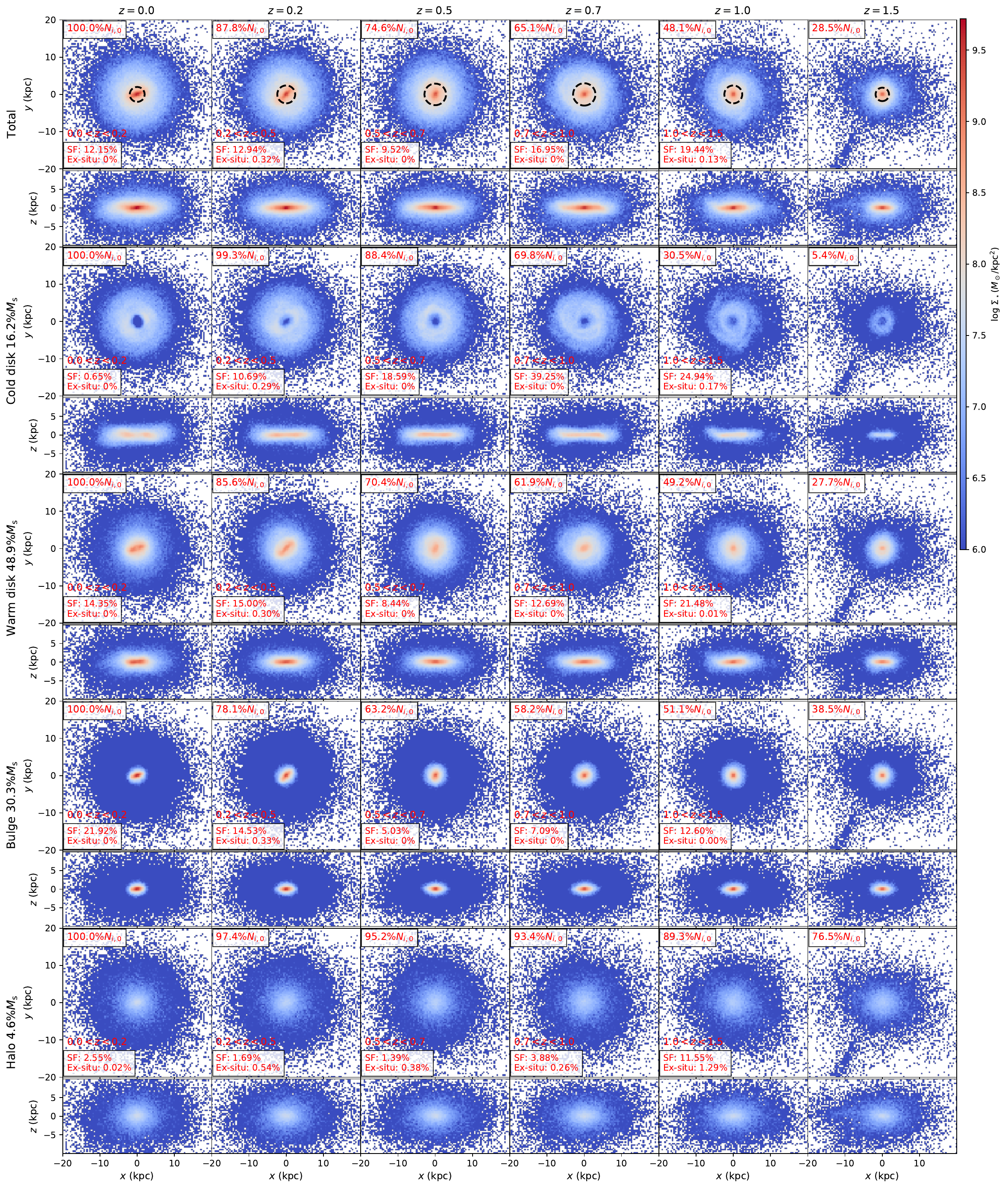}
\caption{B2, a low-mass ($M_{\rm s}\approx 10^{10.2} M_\odot$) bulge-dominated galaxy (ID 590926). This image uses the same convention as \reffig{fig:D1}. The bulge  forms stars until $z=0$, during which $r_{\rm e}$ (marked by the dashed circles) changes little.} 
\label{fig:B2}
\end{center}
\end{figure*}

Because gas seems to be directly accreted into galaxy central regions in B2, it is then
able to contribute directly to the growth of the bulge, shown in \reffig{fig:B2}. B2's bulge keeps growing 
until $z=0$, thus the galaxy size changes little. It is plausible that low-mass bulge-dominated galaxies 
evolve along a similar compact pathway to massive ones, while more massive bulge-dominated galaxies 
evolve naturally either earlier or faster.

\begin{figure}[htbp]
\begin{center}
\includegraphics[width=0.48\textwidth]{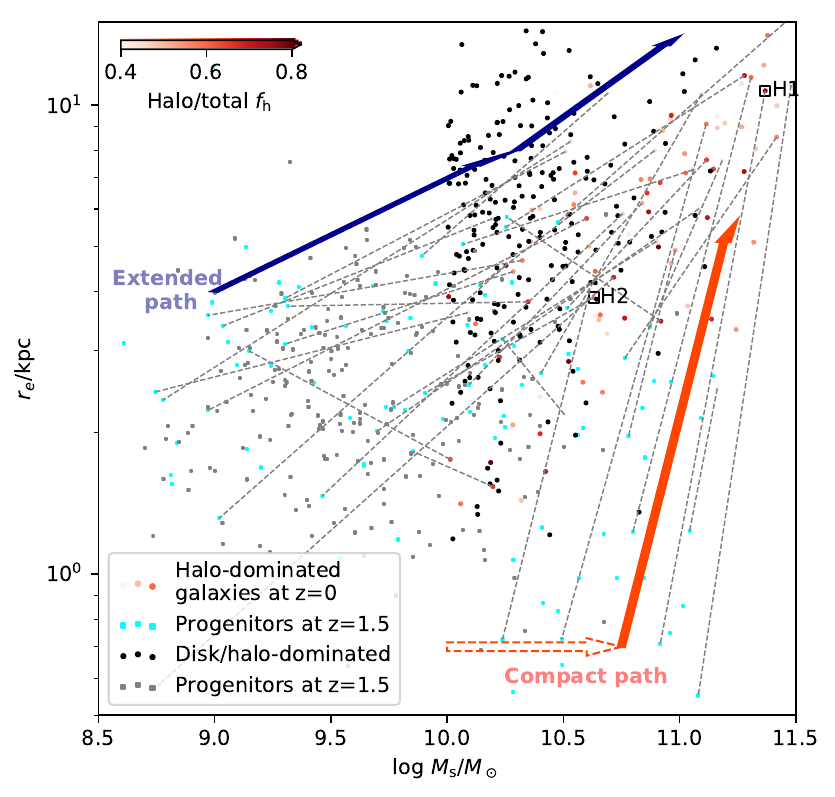}
\caption{Evolution of halo-dominated galaxies (red dots) on the mass-size diagram. The disk/bulge-dominated galaxies (black dots) and their progenitors (gray squares) are overlaid. The cyan squares are the progenitors of halo-dominated galaxies at $z=1.5$. One-third of halo-dominated galaxies are linked with their progenitors using gray dashed lines, which indicate the compact and extended pathways highlighted by red and blue arrows, respectively.}
\label{fig:Hpath}
\end{center}
\end{figure}

\begin{figure}[htbp]
\begin{center}
\includegraphics[width=0.48\textwidth]{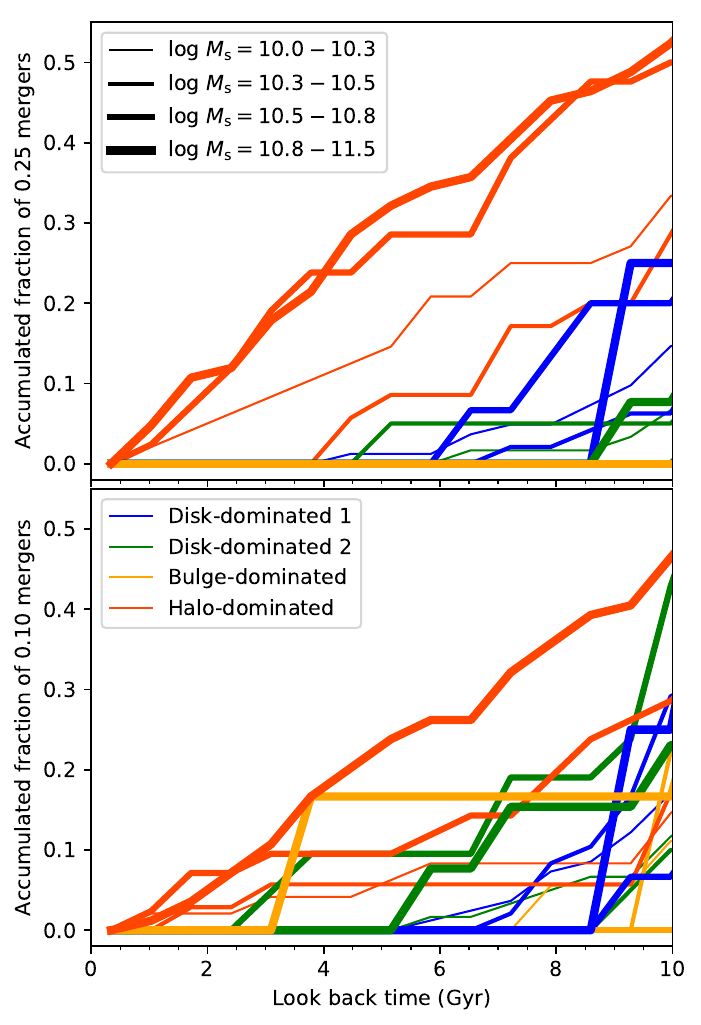}
\caption{Tracing back to 10 Gyr ago, the accumulated fractions of galaxies having major (top, mass ratio of $\geq 0.25$) and minor (bottom, mass ratio of $0.1-0.25$) mergers. For each type of galaxy in a certain mass range, the $y$-axis gives the fraction of galaxies that have been affected by mergers since that time. }
\label{fig:merger}
\end{center}
\end{figure}

\section{Evolution of halo-dominated galaxies: the role of mergers}
\label{sec:halo}

Halo-dominated galaxies generally have weak rotation and elliptical morphology, and thus 
qualitatively correspond to slow rotators in elliptical galaxies. Fast rotators are 
likely to have rotation close to the kinematic criteria of elliptical galaxies, i.e., 
$f_{\rm d}=0.5$ or $K_{\rm rot}=0.5$. In \reffig{fig:Hpath}, we 
show the mass-size diagram of halo-dominated galaxies and their progenitors at $z=1.5$. The 
bulge/disk-dominated galaxies (black dots) and their progenitors (gray squares) are overlaid for 
comparison. It is clear that halo-dominated galaxies preferentially populate the high-mass end, 
while their progenitors are distributed in a broad range of both size and mass. It is natural that 
the progenitors of halo-dominated galaxies are either extended or compact young galaxies before 
mergers happen. The evolution of halo-dominated galaxies, thus, can also be divided into 
extended and compact pathways (highlighted by arrows), as suggested by the gray dashed lines 
that link halo-dominated galaxies and their progenitors at $z=1.5$. 

Halo-dominated galaxies that evolve on the compact pathway originate mainly from compact progenitors that are 
qualitatively similar to those of bulge-dominated galaxies, but generally more massive and compact. 
Such compact objects are likely to be the so-called ``nuggets'' observed in many high redshift observations 
\citep[e.g.][]{vanDokkum2008, vanDokkum2009, Newman2010, Damjanov2011, Whitaker2012, Barro2013}. 
A large fraction of the progenitors of halo-dominated galaxies have similar properties to 
those of disk-dominated galaxies, falling on the extended pathway. 

Mergers are destructive for galaxies along both extended and compact pathways that can disrupt 
galactic spins, thus building stellar halos in the second phase. As shown in \reffig{fig:merger}, 
about 50\% of massive halo-dominated galaxies 
have had at least one $\geq0.25$ major mergers (top panel) in the past 10 Gyr, which is consistent 
with the results of \citet{Penoyre2017} and \citet{Lagos2018}. This fraction is 
smaller ($\sim 30$\%) in less massive ($M_{\rm s} \leq 10^{10.5} M_\odot$) galaxies, suggesting 
that less violent mergers are required to form a halo-dominated galaxy possibly due to their weaker 
potential well. Mergers of mass ratio 0.1-0.25 (bottom panel) play a relatively less important, but  
non-negligible, role. In comparison, less than 20\% bulge-dominated galaxies have been affected by 
any mergers during this time period.

\begin{figure*}[htbp]
\begin{center}
\includegraphics[width=0.95\textwidth]{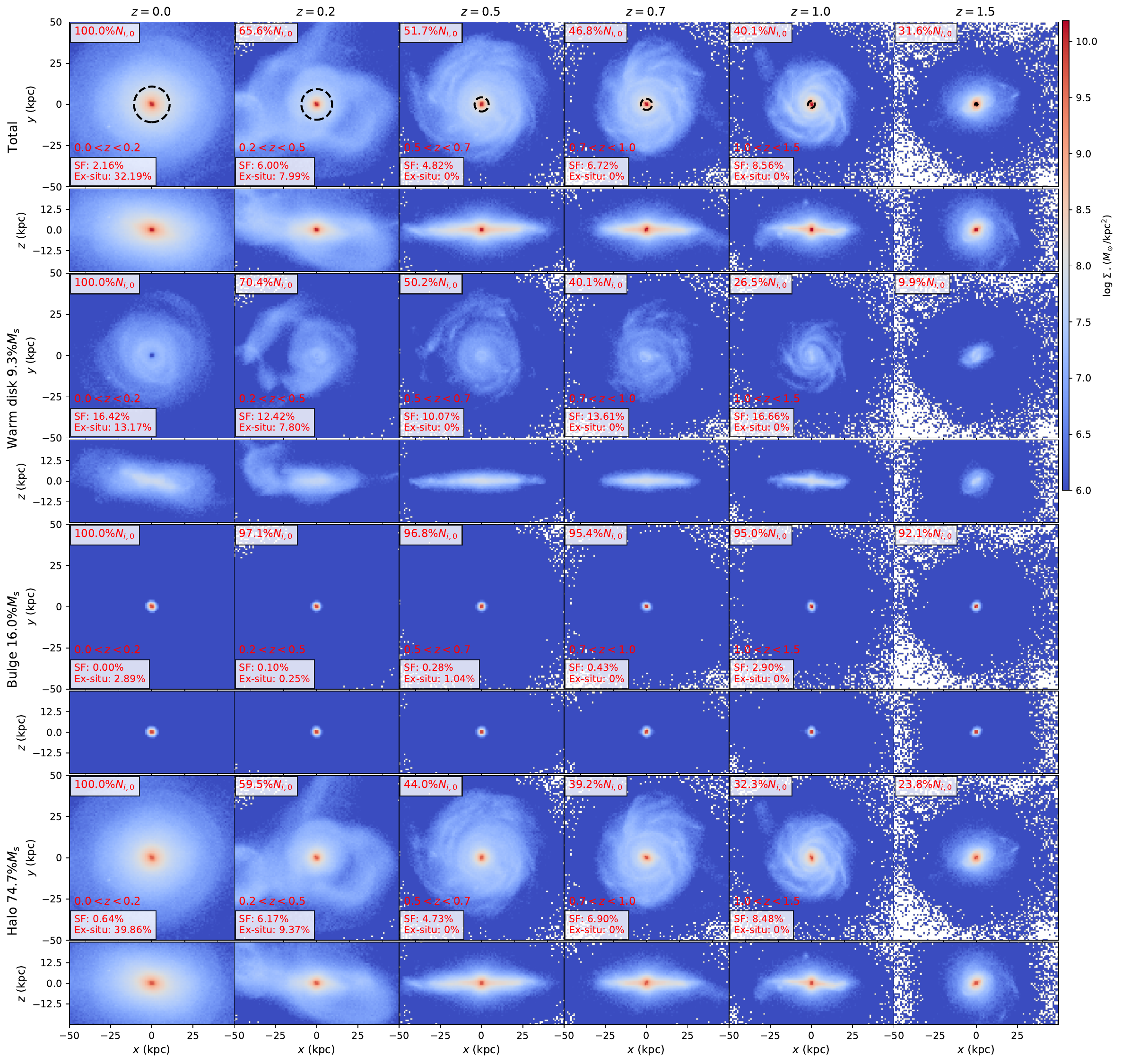}
\caption{H1 (ID 398784), a prototype halo-dominated galaxy evolves along the compact pathway. This massive elliptical galaxy forms by a major merger at $z\sim 0.2$ from a rather compact object at $z=1.5$. The stellar halo is built up by the initial disk of the primary galaxy and the satellite galaxy destroyed during the merger. This image uses the same convention as \reffig{fig:D1}.}
\label{fig:H1}
\end{center}
\end{figure*}

\begin{figure*}[htbp]
\begin{center}
\includegraphics[width=0.95\textwidth]{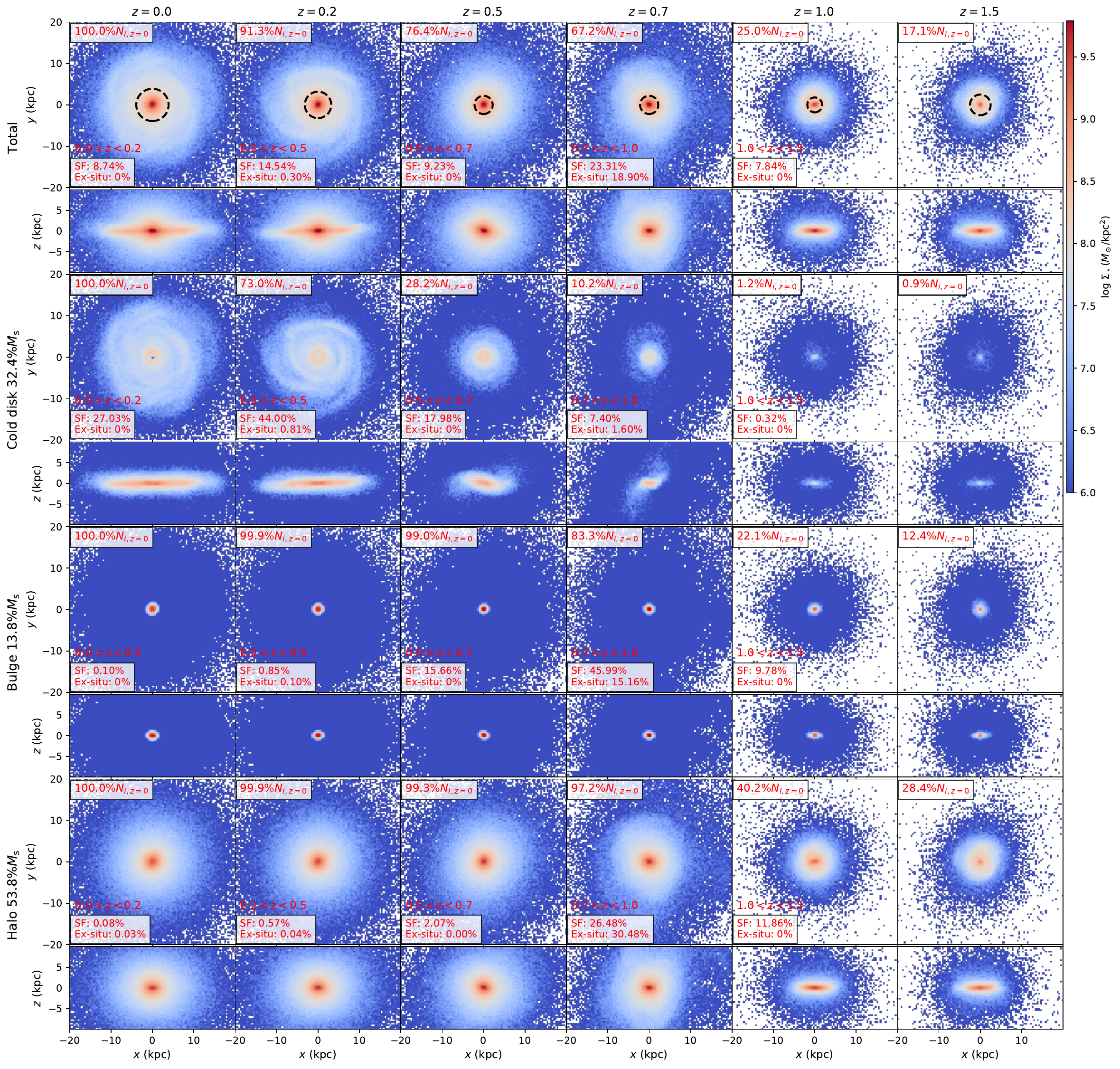}
\caption{A halo-dominated galaxy (H2, ID 559386) evolves along the extended pathway. Its disk is rebuilt after the merger at $z\sim 0.7$, forming a Sombrero analogue.}
\label{fig:H2}
\end{center}
\end{figure*}

\subsection{Two prototypes evolve along the compact and extended pathways}

Two prototype halo-dominated galaxies H1-H2 (marked by squares in \reffig{fig:Hpath}) are 
discussed in this section. They evolve along extended and compact pathways that are shaped 
by mergers, especially major ones.

H1 (\reffig{fig:H1}) is a typical massive elliptical galaxy. At $z=1.5$, the progenitor of this galaxy has 
about $10^{10.6}\ M_\odot$ of stellar mass and a compact morphology of $r_{\rm e} \sim 0.7$ kpc, thus a 
typical nugget. It evolves in a similar way to bulge-dominated galaxies such that a disk is assembled via 
in-situ SF during $z=0.2-1.5$ before the major merger happens at $z\sim 0.2$. If no merger is involved 
in the evolution of this galaxy, it would become a galaxy of $M_{\rm s} \sim 10^{11.0} M_\odot$ and 
$r_{\rm e} \sim 4$ kpc, i.e., a massive bulge-dominated galaxy, according to its properties at $z=0.5$. 
The major merger transforms it to a halo-dominated elliptical galaxy with $M_{\rm s} = 10^{11.3} M_\odot$ and 
$r_{\rm e} \simeq 10$ kpc. The growth of galaxy size is likely a natural outcome combining mergers 
and extended SF via gas accretion. The most dramatic increase of the galaxy size is driven by the 
final major merger, especially for galaxies along the compact pathway in \reffig{fig:Hpath}, 
generating a massive elliptical galaxy with a diffuse envelope. About $50\%$ of the stellar halo particles 
are from the satellite galaxy merging in and SF during the merger event; another $50\%$ come from the 
stars previously existing in the central galaxy. Although the bulge's progenitor of such a galaxy 
is already quite massive, reaching $\sim 10^{10.5}\ M_\odot$, at $z=1.5$, it only contributes to a 
small fraction of the total stellar mass at $z=0$ because of the contribution from ex-situ stars 
accreted during the merger. 

As shown in \reffig{fig:Hpath}, most of halo-dominated galaxies that evolve along the compact pathway become 
more massive by a factor of $\sim 2$ during $z<2$. Meanwhile, their sizes are $\sim 10$ times larger. 
Such an evolutionary pathway is consistent with ``nuggets'' that are believed to eventually become more 
extended elliptical galaxies. In contrast, the stellar masses of the galaxies following the extended pathway 
increase significantly by a factor of $\sim 10$, while their sizes become only a few times larger. Even 
for the most massive cases, there is a non-negligible fraction of halo-dominated galaxies formed from 
the extended pathway. This fraction increases significantly toward the low-mass end. 

\reffig{fig:H2} shows the evolution of a prototype halo-dominated galaxy that forms along the extended pathway. 
It is clear that an initially extended disk is destroyed by a major merger at $z\sim 0.7$, but a new disk with 
$32.4\%\ M_{\rm s}$ is generated in the later time, thus forming a Sombrero analogue. In such galaxies, a 
significant merger can disrupt the secular evolution at certain time, but it cannot shut it down completely. 
We therefore suggest that galaxies cannot be quenched directly by either mergers 
\citep[e.g.][]{Hopkins2006, Hopkins2008} or the growth of halos.

\begin{figure*}[htbp]
\begin{center}
\includegraphics[width=\textwidth]{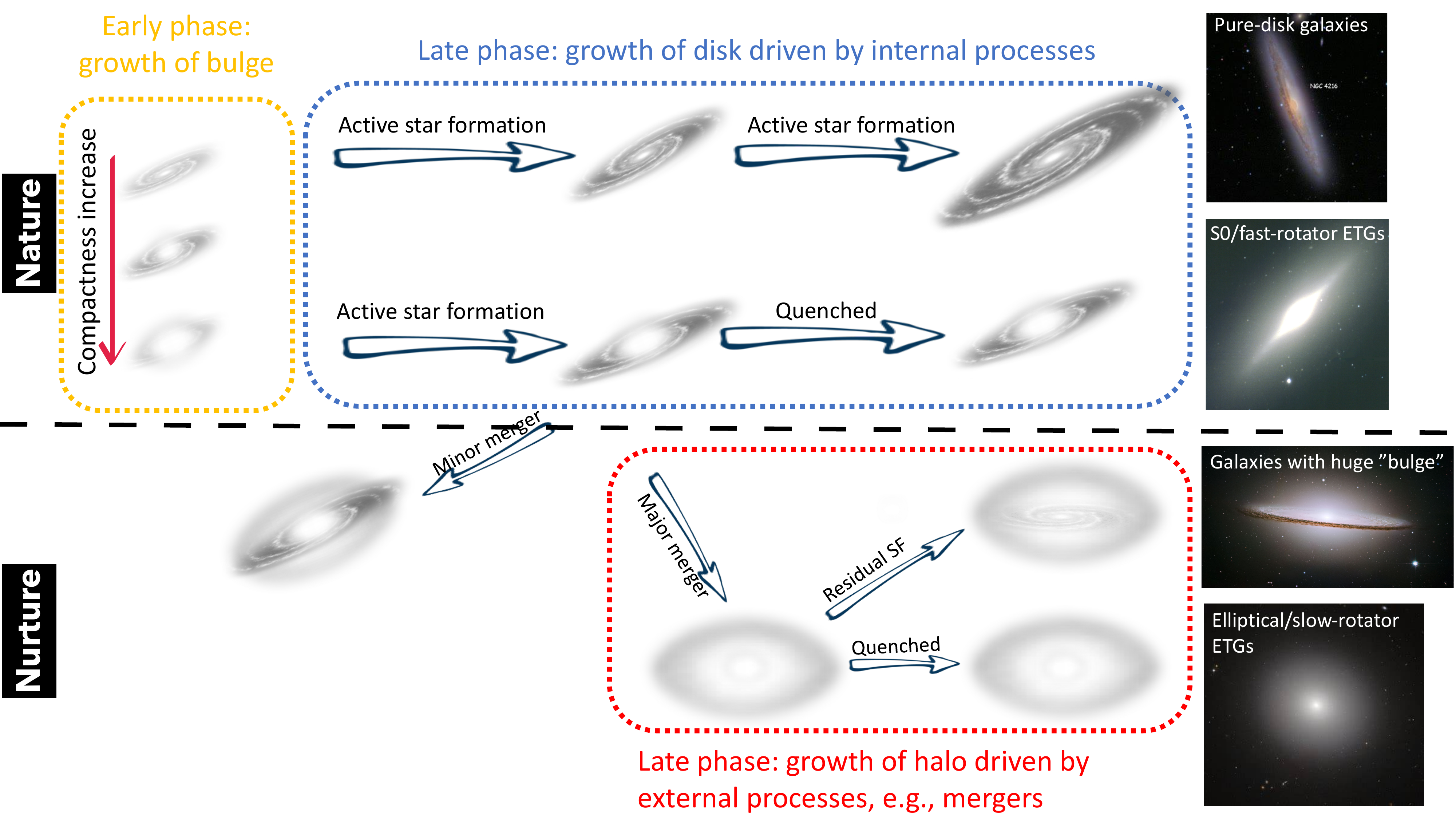}
\caption{Illustration of galaxy evolution suggested by \TNG, based on our kinematic decomposition. The growth of disks, bulges, and halos are physically linked with nature, including early-phase evolution and internal SF, and nurture, mainly merger, processes. Their possible analogues in observations in the Local Universe are suggested in the right panels.}
\label{fig:cartoon}
\end{center}
\end{figure*}

\section{Discussion: bimodality in galaxy types, nature and/or nurture}
\label{sec:discussion}

Galaxies exhibit a bimodality in many aspects, such as colour, SFR, stellar age, and morphology.
Thus, they are generally divided into two main classes: star-forming late-type and quiescent 
early-type galaxies. A similar bimodal distribution of extended SF and compact SF galaxies is also 
found in the Universe early phase in the CANDELS survey \citep{Barro2013, Barro2014, vanDokkum2015}. 
The massive compact galaxies are well know as ``red/blue nuggets'', which are expected to be the 
most likely progenitors of quiescent early-type ones at low redshifts.
In morphology, late- and early-type galaxies are classified by the bulge-disk decomposition, i.e., 
the Hubble (1926) sequence \citep{Sandage1981}. In \citet{Du2020}, we show clearly that the 
morphologically-defined bulge has essentially a severe degeneracy between bulges and stellar halos 
derived by kinematics. In this paper, we further show 
that bulges form from a very different mechanism with respect to stellar halos. This indicates that 
both the nature and nurture should be taken into account to interpret such a bimodal distribution.

\subsection{Bimodality in nature}
\label{sec:nature}
The compact-extended evolutionary pathway can be explained by the long-standing concept 
of the spin parameter \citep{Fall&Efstathiou1980, Blumenthal1984, Mo1998, Dutton2007}. 
The basic idea is that galaxy stellar disks form as a consequence of gas slowly cooling from 
a hot gaseous halo, in the meanwhile, maintaining its specific angular momentum. A remarkable 
scaling relation is found between half-mass radius of the galaxy's stellar distribution and its virial radius, as 
well as the spin, of the galaxy parent halo out to high redshifts 
\citep{Kravtsov2013, vanderWel2014, Shibuya2015, Somerville2018}. Likewise, we also find a 
clear signature that galaxy sizes are partly controlled by halo angular momentum.
Compact galaxies with a more massive bulge generally have a smaller spin than extended 
disk-dominated galaxies at fixed stellar mass, without being affected by mergers. This difference 
finally leads to different evolution along either a compact or extended evolutionary pathway in nature, as 
illustrated in \reffig{fig:cartoon}. Chaotic, 
violent instabilities, i.e., the so-called ``compaction'' phase \citep{Dekel&Burkert2014}, are likely to 
be involved in the compact evolutionary pathway, which may facilitate the growth of bulges.
Moreover, there is a natural downsizing in the compact-extended evolutionary pathway: a compact phase 
occurs earlier in more massive galaxies. At $z=2$, this phase is already over for massive galaxies, forming 
massive compact galaxies, while it is still underway in low-mass ones even at $z=0$.

We suggest that less massive quiescent central galaxies are bulge-dominated 
galaxies that are common in ETGs in the mass range of $10^{10.5} M_\odot < M_{\rm s} < 10^{11} M_\odot$. 
Similarly, \citet{Lagos2018} also reported a quiescent population in less massive galaxies that
have not had any mergers in the EAGLE simulation. Therefore, not all compact galaxies (nuggets) evolve 
into elliptical galaxies. By breaking the degeneracy between bulges and halos, we showed that 
many massive compact galaxies become the bulges in bulge-dominated galaxies, as proposed in 
\citet{Dullo&Graham2013, Graham2013, Graham2015}. A similar conclusion is reached in \citet{Wellons2015, Wellons2016} 
using the Illustris simulation.

\subsection{Bimodality in nurture}
In the later phase, relatively dry mergers, start to be destructive for disk structures in 
galaxies; i.e., the nurture effect in \reffig{fig:cartoon}.
In the general picture, compact galaxies are believed to eventually become more extended 
quiescent galaxies via the cumulative effect of minor mergers that drives the increase of massive quiescent 
galaxies via build-up of a diffuse envelope \citep{Naab2009, Hopkins2010, Oser2010, Porter2014, HuangSong2016}. 
This is partially due to the fact that the number density of quiescent galaxies increases 
by a factor of $\sim 10$ during $z < 2$ in observations \citep{Brammer2011}, which cannot be sufficiently 
explained by the major merger rate during this time \citep{Robaina2010, Brammer2011}. \citet{Genel2018} 
have found a similar trend in the SF and quiescent galaxies from TNG100 that reaches 
a good agreement with observations \citep{Shen2003, vanderWel2014}. Mergers, however, are unable 
to account for the density evolution of less massive quiescent galaxies. Other primary processes, 
e.g., the formation of bulge-dominated galaxies suggested in \refsec{sec:nature}, are 
required to quench the star-forming galaxies in the low-mass end to explain the remaining growth of 
quiescent galaxies since $z\sim 2$. 

In our picture, classical bulges are compact structures formed mainly in the early phase, while 
slow rotator elliptical galaxies are diffuse objects that are dominated by halos formed in the 
later phase. Both the classical bulges and the cores of 
massive elliptical galaxies, however, are likely to be formed in similarly fast SF at high redshifts, 
which is evident by recent observations of red spiral galaxies \citep{HaoCaiNa2019, GuoRui2020, ZhouShuang2020}. 
The difference between the bulge- and halo-dominated galaxies increases in their subsequent evolution, 
largely due to major mergers that lead to a sharp increase in both mass and size of halo-dominated galaxies. 

Therefore, our results suggest that quiescent ETGs are composed of halo- and bulge-dominated galaxies. The bimodality 
in galaxy types is, thus, contributed by both nature and nurture processes. An accurate decomposition of 
bulges and halos is required to understand the formation and evolution of galaxies and their structures.

\section{Summary}
\label{sec:conclusion}

In this work we have studied the origin of galactic stellar structures on the basis of a physically-motivated kinematic decomposition of galaxies from the TNG50 simulation at $z=0$. In particular, we have selected about 500 central galaxies in the $10^{10-11.5}\ M_\odot$ stellar mass range and we have applied the \autoGMM\ method to isolate disks, bulges, and stellar halos in each of them. We have identified three typical kinds of galaxies -- namely, those dominated by disk, bulge, and stellar halo structures, respectively -- and have studied their evolution through cosmic time. 

We find that the growth of structures is characterized and connected by three fundamental regimes: an early-phase evolution ($z\gtrsim2$), followed by late-phase internal processes as well as late-phase external interactions. Our findings motivate an overall framework that can be illustrated as in \reffig{fig:cartoon}.

Galaxies that have massive bulges or disks, but low-mass or negligible stellar halos, have been weakly affected by mergers since $z\sim2$. We find clear indications that bulge- and disk-dominated galaxies evolve along distinct evolutionary pathways, one compact and one extended, respectively, where galaxy sizes are likely to be controlled by the angular momentum obtained by their parent dark matter (proto)halos at early times. In this picture, in the case of low angular momentum, galaxies form stars efficiently in a compact way, by forming bulge-dominated galaxies and by building up massive bulge structures fast in their early-phase evolution. For high angular momentum dark matter halos, disk-dominated galaxies form: stellar disks form as a consequence of gas cooling, during which its specific angular momentum is relatively conserved. In the late phase, both bulge- and disk-dominated galaxies can assemble disky structures that drive the increase of their sizes. This picture suggests that galaxies without diffuse stellar envelopes, i.e., without stellar halo structures, can be used as clean fossil records of their early-phase evolution and properties. 

There is a natural downsizing in the compact-vs.-extended evolutionary picture: more massive galaxies form their bulges earlier. In the case of $M_{\rm s} >10^{10.5}$ at $z=0$, progenitors of bulge-dominated galaxies generally have already been rather massive and compact objects that have similar properties 
to ``nuggets'' observed at high redshifts. This also suggests that some nuggets are likely to become the bulges of massive galaxies in the local Universe. In the later phase, 
such massive bulge-dominated galaxies are quenched inside-out, at least according to TNG50. In less-massive bulge-dominated galaxies, 
their star formation occurs within the disk via gas accretion until recent times. 

Galaxies with massive halos are slow rotator elliptical galaxies whose formation is dominated by major mergers at recent times, i.e. in the later phase. The progenitors of halo-dominated galaxies can therefore be either compact nuggets or extended disk galaxies. Mergers, especially major ones, are able to destroy the stellar disks of galaxies, which in turn contribute to the formation of massive stellar halos. However, mergers alone cannot quench star formation, and disky structures can regenerate also after major mergers. 

In \citet{Du2020}, we showed that stellar halos are significantly mixed up with kinematically-derived bulges, so that it is hard for the two structures to be properly decomposed based on morphology i.e. photometry. In this paper, we have further shown that the inaccurate classification and definition of bulges and stellar halos is destined to cause further difficulties in our understanding of the formation histories of galaxies. This work provides an initial framework for future attempts to link galactic structures to galaxy formation physics in detail.

\begin{acknowledgements}

This work was supported by the National Science Foundation of China (11721303, 11991052) and the National Key R\&D Program of China (2016YFA0400702). The authors thank Vicente Rodriguez-Gomez for his contribution with running {\sc SKIRT} to generate the synthetic images of Figure 5. M.D. is also supported by the grants ``National Postdoctoral Program for Innovative Talents'' (BX201700010) and ``China Postdoctoral Science General Grant'' (2018M630025) funded by China Postdoctoral Science Foundation. V.P.D. was supported by STFC Consolidated grant ST/R000786/1. The TNG50 simulation used in this work, one of the flagship runs of the IllustrisTNG project, has been run on the HazelHen Cray XC40-system at the High Performance Computing Center Stuttgart as part of project GCS-ILLU of the Gauss centres for Supercomputing (GCS). This work is also strongly supported by the High-performance Computing Platform of Peking University, China. The analysis was performed using \texttt{Pynbody} \citep{pynbody}.
\end{acknowledgements}

\appendix

\section{Appendix}
\label{appendix}

\begin{figure*}[htbp]
\begin{center}
\includegraphics[width=1.\textwidth]{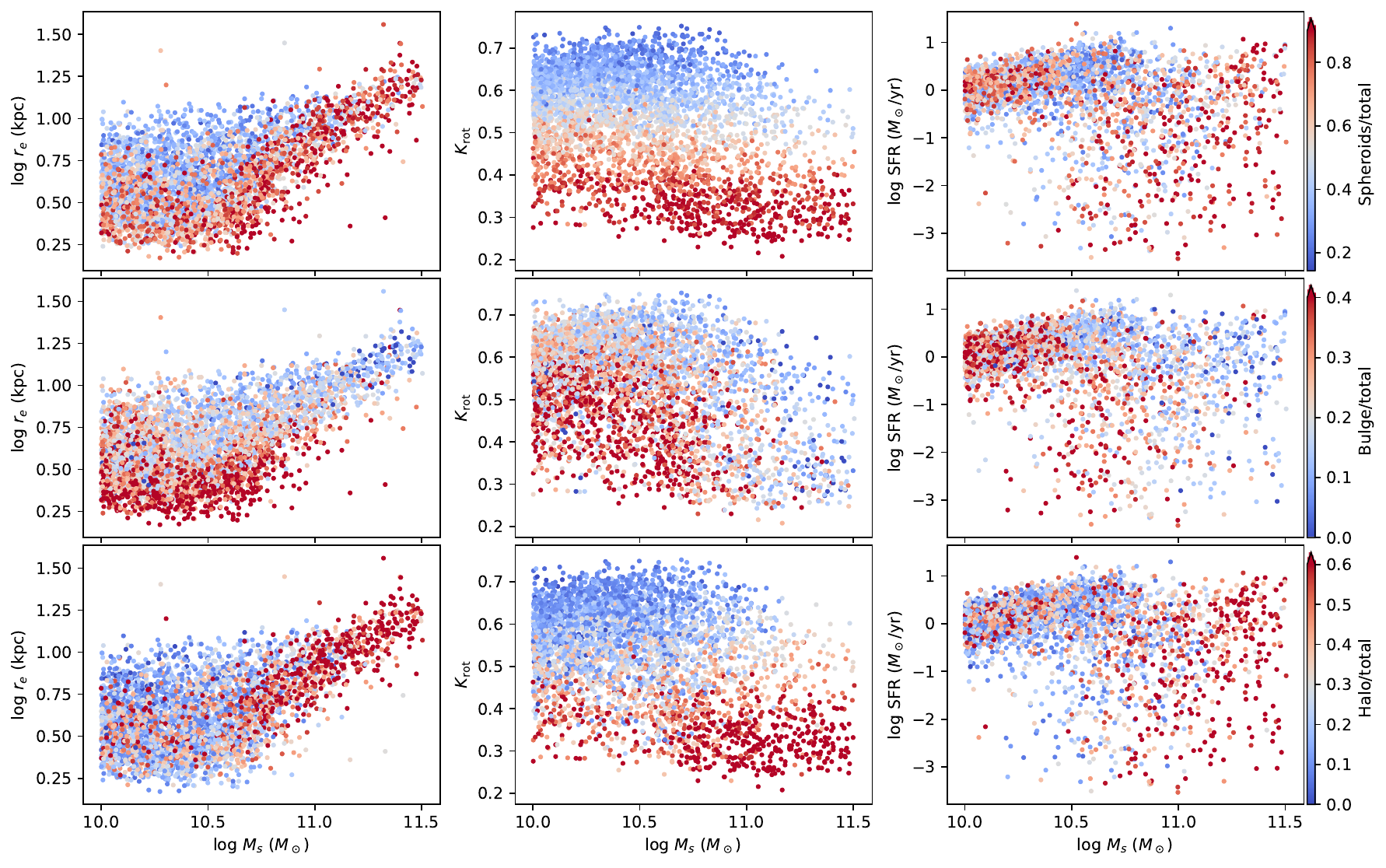}
\caption{Relation between global properties and the mass fractions of kinematically-derived structures for TNG100 central galaxies. This image uses the same convention as \reffig{fig:TNG50basic}. In low-mass galaxies, TNG100 cannot resolve disks well, thus significantly overproducing kinematic bulges.}
\label{fig:TNG100basic}
\end{center}
\end{figure*}

\begin{figure*}[htbp]
\begin{center}
\includegraphics[width=1.\textwidth]{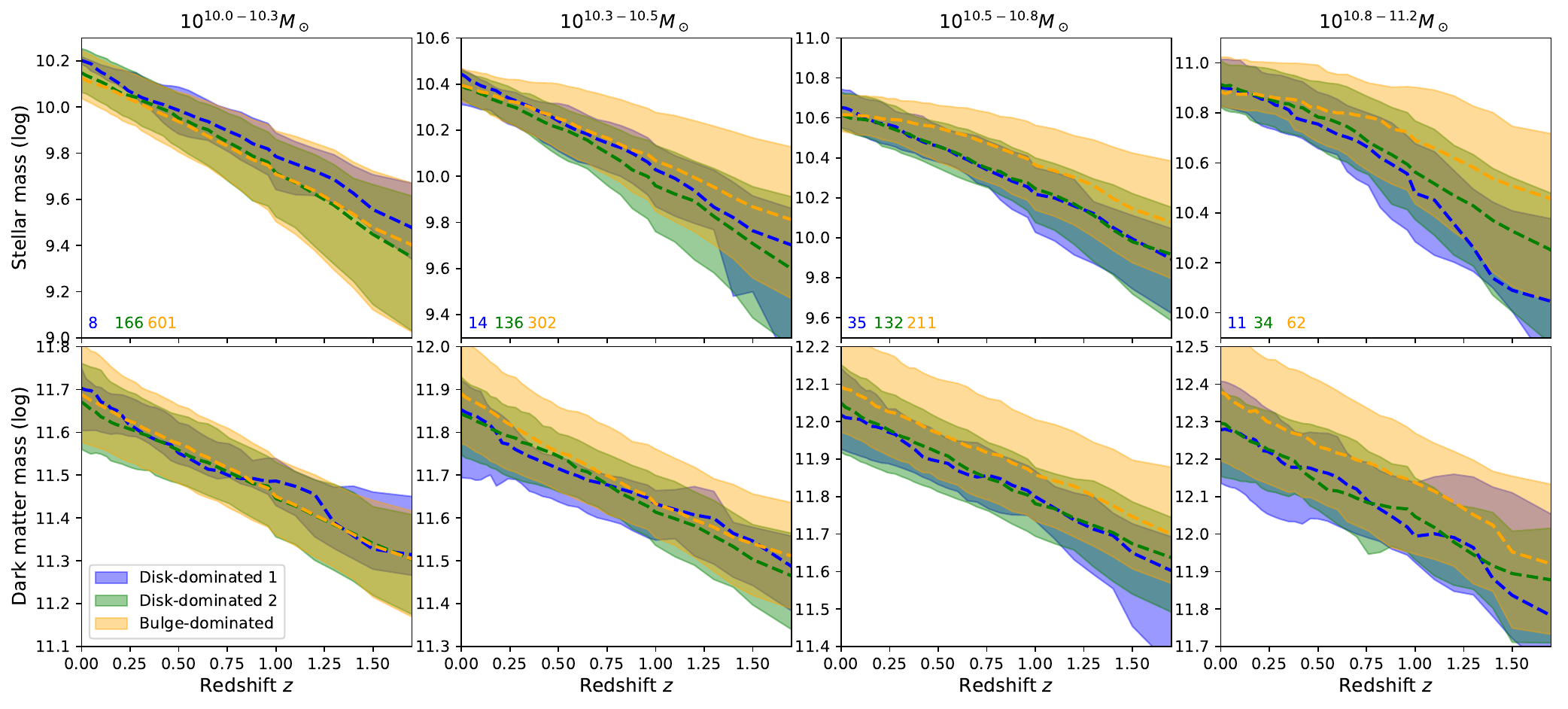}
\includegraphics[width=1.\textwidth]{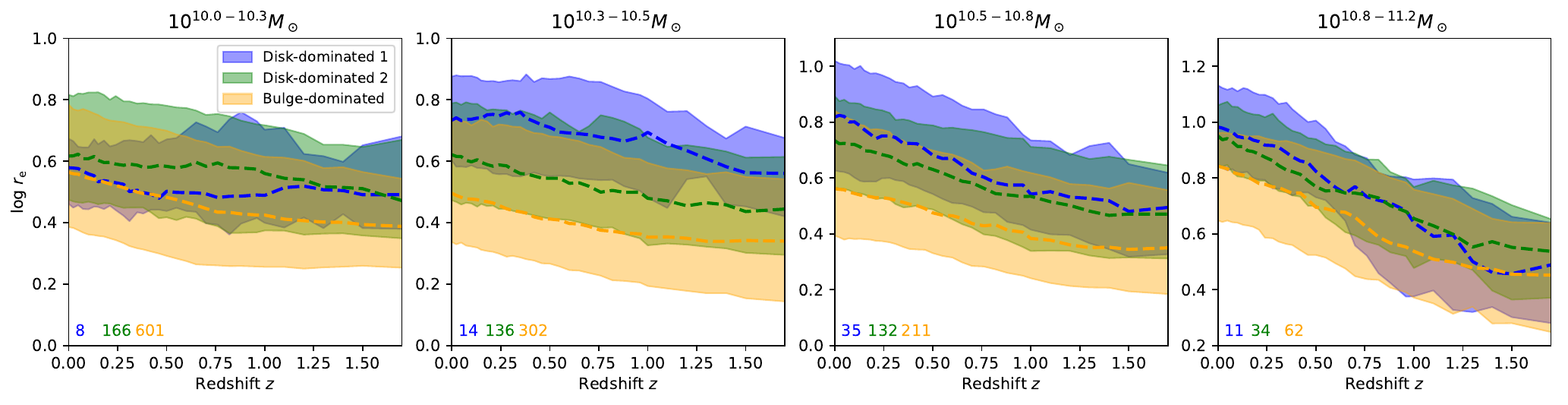}
\includegraphics[width=1.\textwidth]{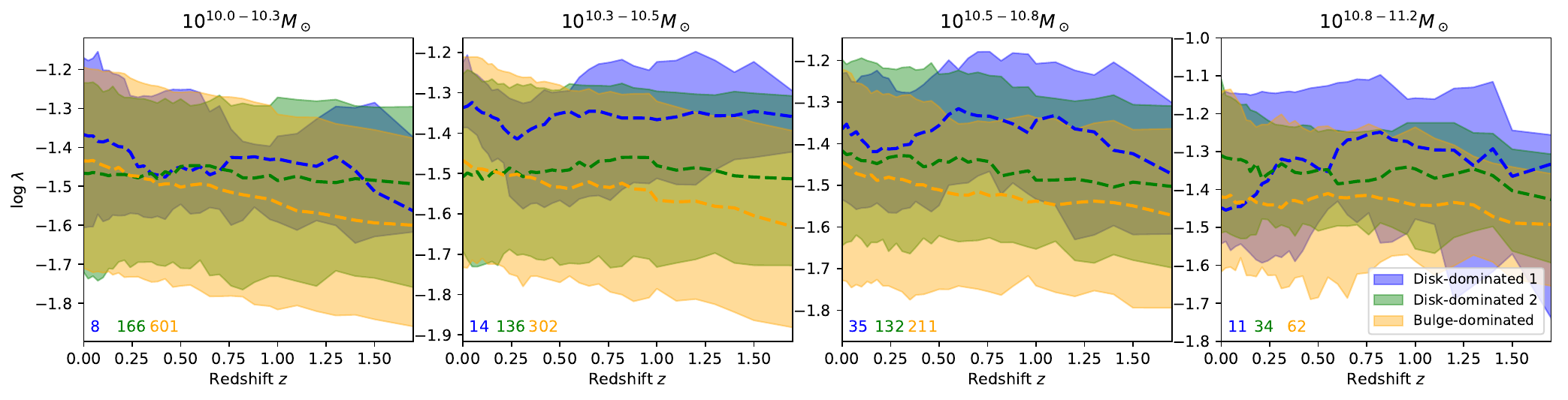}
\caption{From top to bottom: evolution of stellar and dark matter masses, half-mass radius $r_{\rm e}$, and spin parameter for bulge- and disk-dominated galaxies selected in TNG100. This image uses the same convention as \reffig{fig:growth}. Bulge-dominated galaxies generally have lower spins than disk-dominated ones, which may lead to evolution along a more compact evolutionary pathway.}
\label{fig:TNG100evo}
\end{center}
\end{figure*}

For comparison, we perform the same analysis on central galaxies in the same mass 
range from the TNG100 run. \reffig{fig:TNG100basic} shows that the galaxies 
from TNG100 follow a similar trend to those from TNG50 shown in \reffig{fig:TNG50basic}. 
Galaxies with massive spheroidal components are relatively compact and quiescent objects.  
It is worth mentioning that, at the low-mass ($M_{\rm s} \lesssim 10^{10.3} M_\odot$) 
end, many galaxies dominated by spheroidal components are 
still actively forming stars, which is consistent with the blue spheroid issue reported in 
\citet{Rodriguez-Gomez2019}. The increase of bulge mass fraction (\reffig{fig:population}) 
in TNG100 may be due to the overheating of disk stars in central regions where the dynamical time 
is shortest. The difference among different types of galaxies is dramatically weakened 
in this mass range. TNG50 produces much more realistic low-mass galaxies, possibly due to the dramatic 
increase of the numerical resolution that resolves disk thicknesses well \citep{Pillepich2019}. 
\reffig{fig:TNG100evo} shows the evolution of some basic properties of disk- and bulge-dominated 
galaxies from TNG100. We can see a similar compact-extended evolutionary pathways in bulge- and 
disk-dominated galaxies from TNG100.

\bibliography{Reference_lib}

\end{document}